\documentclass[11pt,a4paper]{article}
\pdfoutput=1
\usepackage{jheppub}

\usepackage{booktabs}
\usepackage[english]{babel}
\usepackage{amsmath,amssymb,amsbsy,amstext, amsthm, simplewick}
\usepackage{hyperref}
\usepackage{graphicx}
\usepackage{amsfonts}
\usepackage{amssymb}
\usepackage[small]{caption}
\usepackage{upgreek}
\usepackage{cancel}
\usepackage{color}
\usepackage[dvipsnames]{xcolor}
\usepackage{subcaption}

\usepackage{colortbl}
\definecolor{lightgreen}{cmyk}{0.2, 0, 0.2, 0.2}
\definecolor{lightgray}{cmyk}{0.1,0.2,0,0.1}
\definecolor{lightgray2}{cmyk}{0.1,0.1,0,0.1}

\makeatletter
\newlength{\apb@width}
\newcommand{\autoparbox}[2][c]{\settowidth{\apb@width}{#2}\parbox[#1]{\apb@width}{#2}}

\makeatother

\numberwithin{equation}{section}

\def\beq{\begin{equation}}
\def\eeq{\end{equation}}

\def\bea{\begin{eqnarray}}
\def\eea{\end{eqnarray}}

\def\beq{\begin{equation}}
\def\eeq{\end{equation}}
\def\bea{\begin{eqnarray}}
\def\eea{\end{eqnarray}}

\def\Mp{M_{\rm pl}}

\DeclareRobustCommand{\SkipTocEntry}[4]{}

\newcommand{\mpl}{M_\text{pl}}

\newcommand{\edit}[2]{\ignorespaces}

\newcommand{\appropto}{\mathrel{\vcenter{
  \offinterlineskip\halign{\hfil$##$\cr
    \propto\cr\noalign{\kern2pt}\sim\cr\noalign{\kern-2pt}}}}}

\title{Cosmological signals of a mirror twin Higgs}

\author[\spadesuit,\clubsuit]{Nathaniel Craig,}
\author[\spadesuit]{Seth Koren,}
\author[\spadesuit]{and Timothy Trott}

\affiliation[\spadesuit]{Department of Physics, University of California, Santa Barbara, CA 93106, USA}	
\affiliation[\clubsuit]{Kavli Institute for Theoretical Physics, Santa Barbara, CA 93106, USA}

\emailAdd{ncraig@physics.ucsb.edu}
\emailAdd{koren@physics.ucsb.edu}
\emailAdd{ttrott@physics.ucsb.edu}

\abstract{We investigate the cosmology of the minimal model of neutral naturalness, the mirror Twin Higgs. The softly-broken mirror symmetry relating the Standard Model to its twin counterpart leads to significant dark radiation in tension with BBN and CMB observations. We quantify this tension and illustrate how it can be mitigated in several simple scenarios that alter the relative energy densities of the two sectors while respecting the softly-broken mirror symmetry. In particular, we consider both the out-of-equilibrium decay of a new scalar as well as reheating in a toy model of twinned inflation, Twinflation. In both cases the dilution of energy density in the twin sector does not merely reconcile the existence of a mirror Twin Higgs with cosmological constraints, but predicts contributions to cosmological observables that may be probed in current and future CMB experiments. This raises the prospect of discovering evidence of neutral naturalness through cosmology rather than colliders.}

\arxivnumber{1611.07977}

\begin{document}
\maketitle
\flushbottom

\section{Introduction}

The electroweak hierarchy problem is one of the primary motivators for accessible physics beyond the Standard Model and has led to an expansive set of searches at the LHC and beyond. Recent null results in  searches for conventional approaches to the hierarchy problem motivate the exploration of alternative solutions. ``Neutral naturalness'' provides one such promising alternative, in which the lightest states responsible for protecting the weak scale are partly or wholly neutral under the Standard Model (SM).  In these theories, discrete symmetries enforce cancellations between finite threshold corrections to the Higgs mass. The discrete symmetries may be approximate or exact, although solutions with approximate symmetries typically require a plethora of new particles near the TeV scale.

Perhaps the simplest avatar of neutral naturalness is the ``mirror'' Twin Higgs \cite{Chacko:2005pe}, in which the new physics near the weak scale consists of an identical copy of the Standard Model related by an exact $\mathbb{Z}_2$ exchange symmetry. Higgs portal-type couplings between the Higgs doublets of the Standard Model and the twin sector lead to accidental global symmetries that protect the Higgs mass. The lightest partner particles are entirely neutral under the Standard Model, subject only to indirect bounds from precision Higgs coupling measurements. In conjunction with supersymmetry or compositeness at 5-10 TeV, this provides a complete solution to the ``little'' and ``big'' hierarchy problems consistent with current LHC limits. In this respect, the Twin Higgs naturally reconciles the observation of a light Higgs with the absence of evidence for new physics thus far at the LHC.

The primary challenge to the mirror Twin Higgs comes not from LHC data, but from cosmology. An exact $\mathbb{Z}_2$ exchange symmetry predicts mirror copies of light Standard Model \edit{fermions}{neutrinos and photons} states, which contribute to the energy density of the early universe. In particular, twin neutrinos and a twin photon provide a new source of dark radiation that is strongly constrained by CMB and BBN measurements \cite{Ade:2015xua, Cyburt:2015mya}. While these constraints could be avoided if the two sectors were at radically different temperatures, the Higgs portal couplings required by naturalness keep the two sectors in thermal equilibrium down to relatively low temperatures. Constraints on dark radiation in the mirror Twin Higgs have motivated models in which the $\mathbb{Z}_2$ symmetry is approximate (such as the orbifold \cite{Craig:2014aea, Craig:2014roa}, holographic \cite{Geller:2014kta, Barbieri:2015lqa, Low:2015nqa}, fraternal \cite{Craig:2015pha}, and vector-like \cite{Craig:2016kue} Twin Higgs), in which case the dark radiation component can be made naturally small. This problem was examined recently in \cite{Barbieri:2016zxn}, where the $\mathbb{Z}_2$ symmetry in the fermion Yukawa couplings was broken in order to find an arrangement that would reduce the residual dark radiation from the twin particles.\footnote{For recent related work on the cosmology and cosmological signatures of non-minimal Twin Higgs scenarios, see e.g.~\cite{Garcia:2015loa,Craig:2015xla, Garcia:2015toa, Farina:2015uea, Freytsis:2016dgf,Farina:2016ndq, Prilepina:2016rlq}.} However, such cosmological fixes come at the cost of minimality, as models with approximate $\mathbb{Z}_2$ symmetries require a considerable amount of additional structure near the TeV scale. 

In this work we take an alternative approach and investigate ways in which early universe cosmology can reconcile the mirror Twin Higgs with current CMB and BBN observations. In doing so, we find compelling scenarios that transfer the signatures of electroweak naturalness from high-energy colliders to cosmology. We consider several possibilities in which the energy density of the light particles in the twin sector is diluted by the out-of-equilibrium decay of a new particle after the two sectors have thermally decoupled. Crucially, the new physics in the early universe respects the exact (albeit spontaneously broken) $\mathbb{Z}_2$ exchange symmetry of the mirror Twin Higgs. This symmetry may be used to classify representations of the particle responsible for this dilution. We concentrate on two minimal cases:
In the first, the long-lived particle is $\mathbb{Z}_2$-even and the asymmetry is naturally induced by kinematics. In the second, there is a pair of particles which are exchanged by the $\mathbb{Z}_2$ symmetry and which may be responsible for inflation.\footnote{A third case exists, in which the particle is $\mathbb{Z}_2$-odd. This may additionally be related to the spontaneous $\mathbb{Z}_2$-breaking in the Higgs potential, although we find that a realisation of such a scenario is dependent upon the UV completion of the model.} Moreover, in these cases the new physics does not merely reconcile the existence of a mirror twin sector with cosmological constraints, but predicts contributions to cosmological observables that may be probed in current and future CMB experiments. This raises the prospect of discovering evidence of electroweak naturalness first through cosmology, rather than colliders, and provides natural targets for future cosmological constraints on minimal realizations of neutral naturalness.


This paper is organized as follows: We begin in Section \ref{sec:twin} by reviewing the salient features of the mirror Twin Higgs. In Section \ref{sec:thermal} we discuss the thermal history of the mirror Twin Higgs, with a particular attention to the interactions keeping the Standard Model and twin sector in thermal equilibrium and the cosmological constraints on light degrees of freedom. In Section \ref{sec:late} we present a simple model where the out-of-equilibrium decay of a particle with symmetric couplings to the Standard Model and twin sector leads to a temperature difference between the two sectors after they decouple. We turn to inflation in Section \ref{sec:twinflation}, constructing a model of ``twinflation'' in which the softly broken $\mathbb{Z}_2$-symmetry extends to the inflationary sector and leads to two periods of inflation. The first primarily reheats the twin sector, while the second primarily reheats the Standard Model sector. We conclude in Section \ref{sec:conc}.

\section{The Mirror Twin Higgs}\label{sec:twin}

We begin by briefly reviewing the salient details of the mirror Twin Higgs. The reader is referred to any of the references listed in the previous section for further details. The theory consists of the Standard Model and an identical copy, related by a $\mathbb{Z}_2$ exchange symmetry at a scale $\Lambda \gg v$. The two sectors are connected only by Higgs portal-type interactions between the two $SU(2)$ doublet scalars.\footnote{Here and in what follows we neglect possible kinetic mixing between the two $U(1)_Y$ gauge bosons; such mixing is not generated in the low-energy theory at three loops \cite{Chacko:2005pe}, and may be forbidden in UV completions where the mirror symmetry relates sectors with unified gauge groups.} Subject to conditions on the quartic coupling, the Higgs sector enjoys an approximate $SU(4)$ global symmetry.\footnote{Properly speaking, the model must contain an $SO(8)$ global symmetry in order to enjoy a residual custodial symmetry \cite{Barbieri:2015lqa, Low:2015nqa}, but in linear realizations the $SU(4)$ is sufficient provided that higher-dimensional operators violating the custodial symmetry are adequately suppressed.}

The Higgs potential is best organized in terms of the accidental $SU(4)$ symmetry involving the $SU(2)$ Higgs doublets of the SM and twin sectors, $H_A$ and $H_B$. The general tree-level Twin Higgs potential is given by (see e.g.~\cite{Craig:2015pha})
\bea
V(H_A, H_B) = \lambda (|H_A|^2+|H_B|^2-f^2/2)^2+\kappa(|H_A|^4+|H_B|^4)+\sigma f^2|H_A|^2\label{TwinPot}
\eea
The first term respects the accidental $SU(4)$ global symmetry, the second breaks $SU(4)$ but preserves the $\mathbb{Z}_2$ and the final term softly breaks the $\mathbb{Z}_2$. Clearly, $\kappa,\sigma\ll\lambda$ are required for the $SU(4)$ to be a good symmetry of the potential. The coupling $\kappa$ should naturally be of order the expected $SU(4)$-breaking radiative corrections to the potential induced by Yukawa interactions with the top/twin top, $\kappa\sim 3y_t^4/(8\pi^2)\log(\Lambda/m_{t})\sim 0.1$ for a cut-off $\Lambda\sim 10$ TeV ($y_t$ being the top quark Yukawa coupling and $m_t$ its mass). Requiring $\lambda\gg\kappa$ therefore implies $\lambda\gtrsim 1$. As the SM and twin isospin gauge groups are disjoint subgroups of the $SU(4)$, the spontaneous breaking of the $SU(4)$ coincides with the SM and twin electroweak symmetry breaking. Three Goldstone bosons are eaten by the broken gauge bosons in each sector, leaving one Goldstone remaining. This will acquire mass through the breaking of the $SU(4)$ that is naturally smaller than the twin scale $f$. For future reference, it is convenient to define the real scalar degrees of freedom in the gauge basis as $h_A=\frac{1}{\sqrt{2}}\Re(H_A^0)-v_A$ and $h_B=\frac{1}{\sqrt{2}}\Re(H_B^0)-v_B$, where $\langle H_A^0\rangle =v_A$ and $\langle H_B^0\rangle =v_B$.


The surviving Goldstone boson should be dominantly composed of the $h_A$ gauge eigenstate in order to be SM-like. The soft $\mathbb{Z}_2$-breaking coupling $\sigma$ is required to tune the potential so that the vacuum expectation values (vevs) are asymmetric and that the Goldstone is mostly aligned with the $h_A$ field direction. The (unique) minimum of the Twin Higgs potential (\ref{TwinPot}) occurs at $v_A\approx \frac{f}{2}\sqrt{\frac{\lambda(\kappa-\sigma)-\kappa\sigma}{\lambda\kappa}}$ and $v_B\approx \frac{f}{2}\sqrt{\frac{\sigma+\kappa}{\kappa}}$. The required alignment of the vacuum in the $H_B$ direction occurs if $\sigma\approx\kappa$, which has been assumed in these expressions for the minimum. The consequences of this are that $v_A\approx v/\sqrt{2}$ and $v_B\approx f/\sqrt{2}\gg v$ (where $v$ is the vev of the SM Higgs, although $v_A\approx 174$ GeV is the vev that determines the SM particle masses and electroweak properties), so that the SM-like Higgs $h$ is identified with the Goldstone mode and is naturally lighter than the other remaining real scalar, a radial mode $H$ whose mass is set by the scale $f$. The component of $h$ in the $h_B$ gauge eigenstate is $\delta_{hB}\approx v/f$ (to lowest order in $v/f$). Measurements of the Higgs couplings restrict $f\gtrsim 3v$ \cite{Craig:2015pha}, and the naive tuning of the weak scale associated with this inequality is of order $f^2 / 2 v^2$. 

The spectrum of states in the broken phase consists of a SM-like pseudo-Goldstone Higgs $h$ of mass $m_h^2 \sim 8 \kappa v^2$, a radial twin Higgs mode $H$ of mass $m_H^2 \sim 2 \lambda f^2$, a conventional Standard Model sector of gauge bosons and fermions and a corresponding mirror sector. The current masses of quarks, gauge bosons, and charged leptons in the twin sector are larger than their Standard Model counterparts by $\sim f/v$, while the twin QCD scale is larger by a factor $\sim \left(1 +\log(f/v)\right)$ due to the impact of the higher mass scale of heavy twin quarks on the renormalisation group (RG) evolution of the twin strong coupling. The relative mass of twin neutrinos depends on the origin of neutrino masses, some possibilities being $\sim f/v$ for Dirac masses and $\sim f^2/v^2$ for Majorana masses from the Weinberg operator. Mixing in the scalar sector implies that the SM-like Higgs couples to twin sector matter with an $\mathcal{O}(v/f)$ mixing angle, as does the radial twin Higgs mode to Standard Model matter. These mixings provide the primary portal between the Standard Model and twin sectors.

The Goldstone Higgs is protected from radiative corrections from $\mathbb{Z}_2$-symmetric physics above the scale $f$. While the mirror Twin Higgs addresses the little hierarchy problem, it does not address the big hierarchy problem, as nothing stabilizes the scale $f$ against radiative corrections. However, the scale $f$ can be stabilized by supersymmetry, compositeness, or perhaps additional copies of the twin mechanism without requiring new states beneath the TeV scale. Minimal supersymmetric UV completions can furthermore remain perturbative up to the GUT scale \cite{Chang:2006ra}, \cite{Craig:2013fga}.

\section{Thermal History of the Mirror Twin} \label{sec:thermal}

The primary challenge to the mirror Twin Higgs comes from cosmology, rather than collider physics. The mirror Twin contains not only states responsible for protecting the Higgs against radiative corrections (such as the twin top), but also a plethora of extra states due to the $\mathbb{Z}_2$ symmetry that are irrelevant to naturalness. The lightest of these, namely the twin photon and twin neutrinos, contribute significantly to the energy density of the early universe around the era of matter-radiation equality, since they have a temperature comparable to that of the Standard Model plasma at all times. This is because the same Higgs portal coupling that makes the Higgs natural also keeps the two sectors in thermal equilibrium down to $\mathcal{O}$(GeV) temperatures. Then the identical particle content in the twin and Standard Model sectors guarantees that they remain at comparable temperatures even after they decouple - for every massive Standard Model species that becomes non-relativistic and transfers its entropy to the rest of the plasma, its twin counterpart does the same within a factor of $f/v$ in temperature.

In this section we undertake a detailed study of the decoupling between the Standard Model and twin sectors as well as the constraints from precision cosmology.

\subsection{Twin Degrees of Freedom} \label{Sec:Limits}

\begin{figure}[h!]
\centering
\includegraphics[width=1.0\linewidth]{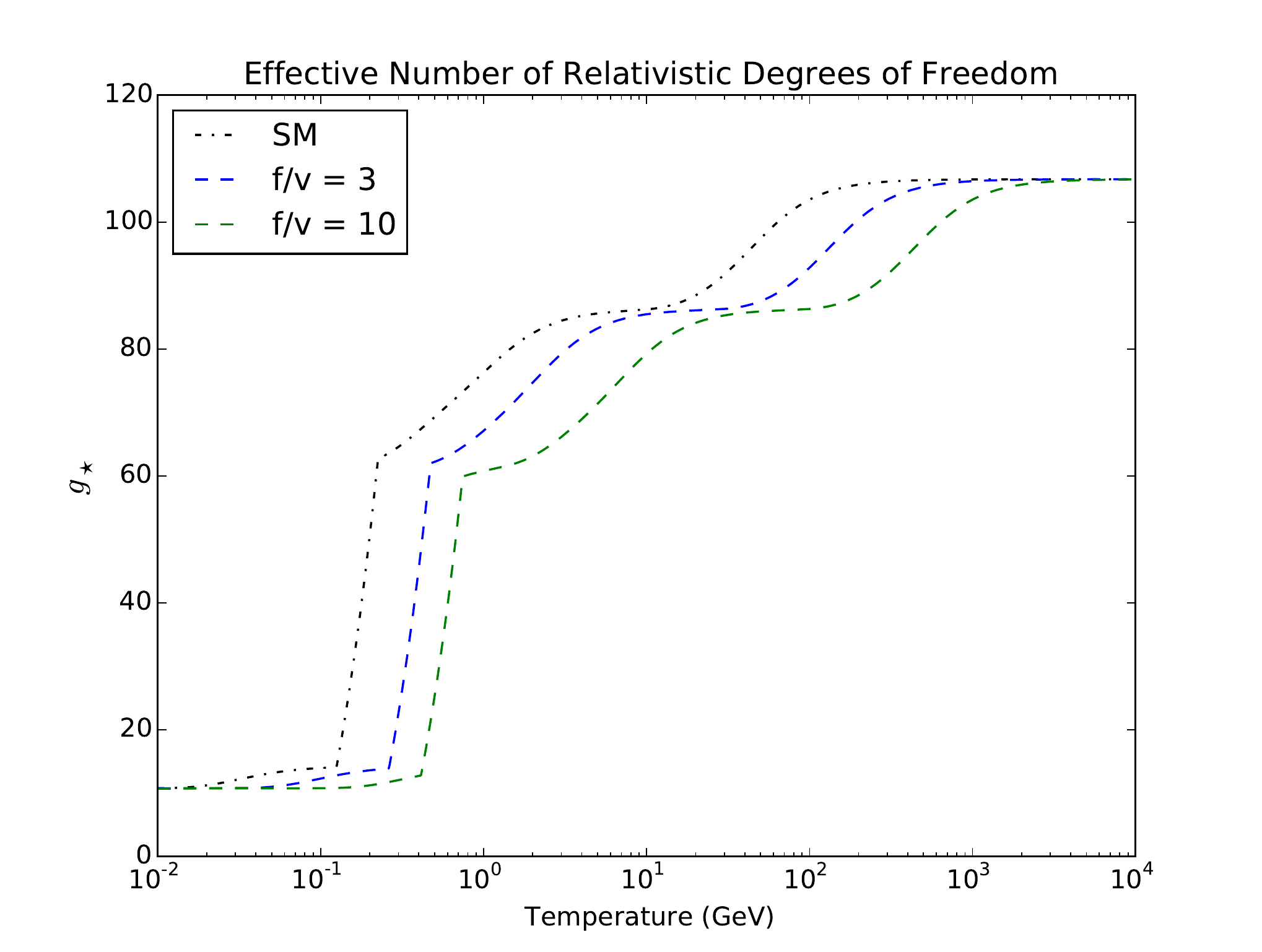}
\caption{The effective number of relativistic degrees of freedom for mirror Twin Higgs models for different values of $f/v$. The dash-dotted line is the for the Standard Model $g^{\rm SM}_\star(T)$ and the dashed lines are the twin sector degrees of freedom $g^{\rm t}_\star(T)$. 
The evolution of $g_\star$ during the QCD phase transition (QCDPT) is not well-understood, so we assign the SM QCDPT a central value of $175$ MeV and a width of $50$ MeV and interpolate linearly between the values of $g_\star$ at $225$ MeV for free partons and at $125$ MeV for pions. Further discussion may be found in \cite{Drees:2015exa}. For the twin sector we use a central value and width which are $(1 + \log(\frac{f}{v}))$ times larger than the SM values. Note that new mass thresholds, expected to appear at energies $\sim 10$ TeV in UV completions of the twin Higgs, have not been included.}
\label{Fig:geff}
\end{figure}

In thermal equilibrium, each relativistic degree of freedom has roughly the same energy density. In general, we express the energy density of the universe $\rho$ during the radiation-dominated era as $\rho \equiv g_{\star} \frac{\pi^2}{30} T^4,$ where we define $g_\star$ through this relation as the effective number of relativistic degrees of freedom and $T$ the temperature of the SM photons. This then determines the evolution of the scale factor through the first Friedmann equation
\bea
H = \frac{1}{\mpl} \left[ \frac{\pi^2}{90} g_\star T^4\right]^{1/2} 
\eea
(assuming spatial flatness), where $\mpl$ is the reduced Planck mass. In general, the energy density of a particular species $i$ may be computed from $\rho_i = g_i \int \frac{d^3p}{(2\pi)^3} f_i(p,T_i) E(p)$, where $g_i$ are the number of internal degrees of freedom, $E(p)$ is the energy as a function of momentum $p$, while $f_i(p,T_i)$ is the phase-space number density and is a Bose-Einstein or Fermi-Dirac distribution if the species is in equilibrium at temperature $T_i$. The number of effective relativistic degrees of freedom may then be defined for each sector separately as $g_{\star}^{\text{SM}}(T)$ and $g_{\star}^{\text{t}}(T) $ satisfying $\rho_{\text{SM}}(T)=\frac{\pi^2}{30}g_{\star}^{\text{SM}}(T)  T^4$ and $\rho_{\text{t}}(T)=\frac{\pi^2}{30}g_{\star}^{\text{t}}(T) T^4$, respectively, where $\rho_{\text{SM}}(T)$ and $\rho_{\text{t}}(T)$ are the total energy densities of SM and twin particles. The values of $g_{\star}(T)$ for the SM and twin sectors are shown in Figure \ref{Fig:geff}, where all species within each sector are in thermal equilibrium. These can then be used to calculate the total number $g_\star$ as a function of temperature, by weighting twin sector energy density by its temperature: $g_\star(T)=g_\star^{\text{SM}}(T)+g_\star^{\text{t}}(\hat{T})(\hat{T}/T)^4$, where $\hat{T}$ is the twin sector photon temperature when the SM photon temperature is $T$.

Likewise, entropy densities for each sector $i$ are defined as $s_i(T)=\frac{2\pi^2}{45}g^{\text{i}}_\star(T) T^3$. We neglect the small differences between the number of relativistic degrees of freedom defined from energy and entropy densities, which are not significant over the range of temperatures of interest here.

\subsection{Decoupling} \label{Sec:Decoupling}


In the early universe, the two sectors are thermally linked by interactions mediated by the Higgs, which, through mixing with both $h_A$ and $h_B$ components, allows for SM fermions and weak bosons to scatter off or annihilate into their twin counterparts. However, once the temperature drops sufficiently for this Higgs-mediated interaction to become rare on the expansion time-scale, the sectors decouple and thereafter thermally evolve independently. More precisely, thermal decoupling will occur once the rate at which energy can be exchanged between SM and twin particles (through the Higgs) falls below the Hubble rate. 

Thermal decoupling is traditionally formulated from the Boltzmann equations describing the evolution of single-particle phase space number densities, wherein collisions induce instantaneous changes to the shape of these distributions. When the collisions occur faster than the expansion rate, the phase space probability density functions of the interacting species are expected to relax to an equilibrium distribution (Boltzmann, neglecting quantum statistics, will be applicable to our case). However, once the rate of collisions falls below the expansion rate, collisions become rare on cosmological time scales and the phase space distributions depart from equilibrium. The decoupling temperature is determined as that at which the scattering rate of a participating particle, $\Gamma$, drops below the Hubble rate, assuming that this occurs instantaneously across the entire phase space where the number density is significant. This formulation can be used to determine the time at which a particular species of particle will cease to scatter off twin particles on cosmological time scales.

In the case of interest here, however, both sectors of particles remain thermalised within themselves while the interactions between sectors freeze-out. This implies that the phase space number densities are still Boltzmann distributions throughout decoupling, with a different temperature for each sector. As it is the twin sector temperature that ultimately determines the impact of the light twin degrees of freedom on the cosmological observables (discussed below in Section \ref{Sec:Limits}), we wish to describe the thermal evolution of the two sectors by that of their entire energy or entropy content and the bulk heat flows between them. They may then be identified as thermally decoupled once the rate at which they exchange energy falls below the expansion rate. 


If the SM and twin sector plasmas have temperatures $T$ and $\hat{T}$ respectively, then calling $q$ the net heat flow density from the SM to the twin sector, the rate at which the twin entropy densities $s_{\text{t}}$ and $s_{\rm SM}$ evolve is determined by
\bea
\frac{ds_{\text{t}}}{dt}+3Hs_{\text{t}}&=&\frac{1}{\hat{T}}\frac{dq}{dt}=\frac{1}{\hat{T}}\Big(\frac{dq_{\rm in}}{dt}-\frac{dq_{\rm out}}{dt}\Big)\label{ThermEv}\\
\frac{ds_{\text{SM}}}{dt}+3Hs_{\text{SM}}&=&\frac{-1}{T}\frac{dq}{dt}=-\frac{1}{T}\Big(\frac{dq_{\rm in}}{dt}-\frac{dq_{\rm out}}{dt}\Big).
\eea
Here, $H$ is the Hubble rate. The heat flow rate has been decomposed into the sum of the energy transferred into and out of the twin sector by collisions in the second equality in each line, where $\frac{dq_{\rm in}}{dt}$ and $\frac{dq_{\rm out}}{dt}$ are both positive.

The rate of heat flow $q$ may be calculated by performing a phase space average of the rate that energy is transferred from the SM to the twin sector through particle interactions. Since the decay rates of top quarks or weak bosons are fast compared to their scattering rate and the Hubble rate, energy transferred to them is instantaneously transferred to the rest of the plasma. Similarly, the scattering rate of lighter fermions off other particles of the same sector (such as photons or gluons) is much faster than their interaction rate with twin fermions. Energy transferred to the lighter fermions therefore quickly diffuses throughout their respective plasmas. The rate of heat flow between sectors may therefore be well approximated by the rate at which energy is transferred from SM particles to twin particles in Higgs mediated interactions. This may occur through elastic scattering of SM particles off twin particles or annihilations of SM particle/antiparticle pairs into twin particles (or the reverse). The energy density transferred to twin particle $i$ from SM particle $j$ in scattering is given by 
\bea
\frac{dq_{ij\rightarrow ij}}{dt}=\frac{g_ig_j}{(2\pi)^6}\int\int \frac{d^3k}{2E_i(k)} \frac{d^3h}{2E_j(h)}f_i(k,\hat{T})f_j(h,T)\,\Big(4E_i(k)E_j(h)\int v_{rel}(E_i(p)-E_i(k))\frac{d\sigma_{ij\rightarrow ij}}{d\Omega}d\Omega\Big), \hspace{2mm} \label{energytrans}
\eea
where $p$ is the outgoing 4-momentum of particle $i$. In the cosmic comoving frame, the phase space number densities $f_i$ and $f_j$ are just Boltzmann factors, although evaluated at the different temperatures of each sector. The factor $g_i$ is the number of internal degrees of freedom of particle $i$, which here includes colour (the cross section should not be colour averaged, as each colour of quark is present in the plasma in equal abundances and each mediates the exchange of energy, so have their contributions summed). Finally, $E_i(k)$ is the on-shell energy of particle $i$ with momentum $k$, while $\frac{d\sigma_{ij\rightarrow ij}}{d\Omega}$ is the differential scattering cross section for species $i$ scattering off $j$ per solid angle $\Omega$ and $v_{rel}$ is the usual relative speed of the incoming particles. As described in \cite{Kelly:1969}, the factor in the integrand giving the energy transferred per reaction is simply a component of a 4-vector, 
\bea
X=4E_i(k)E_j(h)\int (p-k)v_{rel}\frac{d\sigma_{ij\rightarrow ij}}{d\Omega}d\Omega.
\eea
This may be calculated in the centre-of-mass frame and then boosted back into the cosmic comoving frame where the integrals in (\ref{energytrans}) can be evaluated, similarly to the thermal averaging procedure described in \cite{Edsjo:1997bg}. 

The integral (\ref{energytrans}) may be decomposed into two terms giving the positive and negative energy changes of the twin particle, which respectively contribute to $\frac{dq_{\rm in}}{dt}$ and $\frac{dq_{\rm out}}{dt}$. When evaluated in the centre-of-mass frame, these terms correspond to the cases where the scattering angle of the twin particle is respectively less than and greater than the angle between its initial momentum and the total momentum of the system. However, when $T\neq\hat{T}$, we find the integrals involved in this decomposition substantially more arduous than when they are evaluated together.

Energy transferred through annihilations may be similarly calculated as 
\bea
\frac{dq_{j\bar{j}\rightarrow i\bar{i}}}{dt}=\frac{g_j^2}{(2\pi)^6}\int\int \frac{d^3k}{2E_j(k)} \frac{d^3h}{2E_j(h)}f_j(k)f_j(h)\,\Big(4E_j(k)E_j(h)\int v_{rel}(E_j(h)+E_j(k))\frac{d\sigma_{j\bar{j}\rightarrow i\bar{i}}}{d\Omega}d\Omega\Big)\nonumber\\
-\frac{g_i^2}{(2\pi)^6}\int\int \frac{d^3k}{2E_i(k)} \frac{d^3h}{2E_i(h)}f_i(k)f_i(h)\,\Big(4E_i(k)E_i(h)\int v_{rel}(E_i(h)+E_i(k))\frac{d\sigma_{i\bar{i}\rightarrow j\bar{j}}}{d\Omega}d\Omega\Big), \hspace{2mm} \label{HeatFlow}
\eea
where $\frac{d\sigma_{j\bar{j}\rightarrow i\bar{i}}}{d\Omega}$ is now the differential annihilation cross section. This rate may be evaluated as described above and is more directly amenable to the factorisation of the integrals observed in \cite{Edsjo:1997bg}. See also \cite{Adshead:2016xxj} for further details of similar calculations. The first term of (\ref{HeatFlow}) is the energy transferred from the SM to the twin sector and contributes to $\frac{dq_{\rm in}}{dt}$ in (\ref{ThermEv}), while the second term is the energy transferred from the twin sector to the SM and contributes to $\frac{dq_{\rm out}}{dt}$.

In thermal equilibrium, the rate of energy transferred through collisions into one sector will be balanced by that of energy transferred out of it so that there is negligible net heat flow. This state will be rapidly attained (compared to the age of the universe) if $\frac{d q_{\rm in,out}}{dt}\gg 3H\hat{T}s_{\rm t}$. However, as the universe expands and the plasma cools, the energy transfer rates fall faster than the Hubble rate. This is demonstrated in the Figure \ref{Fig:decoupling} below. 
Once they drop below the Hubble rate, energy exchange ceases on cosmological time scales and the sectors thermally decouple, thereafter thermodynamically evolving independently.





To determine the decoupling temperature of the sectors, we calculate the rates of positive energy exchange for the twin particles interacting with the SM particles. The cross sections are calculated using a tree-level effective fermion-twin fermion contact interaction that, in the full twin Higgs model, would be UV completed by a SM Higgs exchange (the heavier mass of the radial mode would make its exchange subdominant). The interaction strength is determined by the masses of the fermions through their Yukawa couplings, as well as the mixing angle of the SM-like mass state $h$ with the gauge eigenstate $h_B$, giving a 4-fermion coupling of strength $\frac{m_f m_{\hat{f}}}{m_h^2f^2}$ (here $m_f$ and $m_{\hat{f}}$ are the masses of fermions $f$ and $\hat{f}$). See \cite{Chang:2006ra}, \cite{Barbieri:2016zxn} for a more detailed discussion of the cross sections. This effective interaction is appropriate for the temperatures of interest here and helps to simplify the integrals of (\ref{energytrans}). In order to further simplify the integrations of (\ref{energytrans}) when it is to be decomposed into terms in which the energy exchange is positive and negative, we calculate $\frac{dq_{\rm in}}{dt}$ under the assumption that the sectors have the same temperature (this ensures that the rate $\frac{dq_{\rm out}}{dt}$ is identical). This is then combined with the rate of energy transferred from annihilation. A similar calculation of these rates was recently performed in \cite{Barbieri:2016zxn}, for cases where the Yukawa couplings do not respect the $\mathbb{Z}_2$ twin symmetry. 

In Figure \ref{Fig:decoupling} we compare the energy transfer rate to the Hubble rate in order to determine when decoupling occurs. As long as the energy exchange rate exceeds the expansion rate, the sectors will be thermalised and have the same temperature. Decoupling then occurs once this rate drops below the Hubble rate. From Figure \ref{Fig:decoupling}, this occurs at a temperature $\sim 2$ GeV. However, even after the energy exchange rate drops below the Hubble rate, the sectors will remain at the same temperature unless some event that either injects or redistributes entropy occurs within a sector (such as the temperature dropping below a mass threshold). As the heavy quark masses roughly coincide with the decoupling temperature, these do cause the twin sector to be mildly reheated with respect to the SM below decoupling. However, the resulting temperature difference is small and the energy exchange rates are expected to continue to be well-approximated by the rates presented in Figure \ref{Fig:decoupling} beyond decoupling.

The lower plot of Figure \ref{Fig:decoupling} illustrates the decomposition of the energy exchange rates into contributions from interactions involving different SM quarks. The interaction cross sections are proportional to the Yukawa couplings of the interacting fermions. The greatest heat exchange is therefore expected to be mediated by the most massive particles, provided that their abundances are not too Boltzmann suppressed. As expected, at temperatures $\sim 1$ GeV, the bottom quark is the best conduit of thermal equilibration, followed by the charm quark and then the $\tau$ (with colour factors enhancing the former two with respect to the latter). The rate of heat flow that the top quarks and weak bosons can mediate at these temperatures (or below) is negligible because of Boltzmann suppression. The bend in the curves at temperatures $\sim 5$ GeV in the lower plot corresponds to a transition from temperatures where the dominant energy exchange rate is through scatterings to those where it occurs through annihilations, as can be seen in the upper plot. The annihilation rate into twin bottom quarks is the dominant component at high enough energies (again because of the larger Yukawa coupling), but this becomes rapidly threshold suppressed as the temperature drops. As can also be inferred in the upper plot, the energy exchange rate through annihilations involving the twin charmed quarks and tau leptons overtakes that of twin bottom quarks at similar temperature, but are still subdominant to scatterings.

\begin{figure}[h!]
\centering
\includegraphics[width=.8\textwidth]{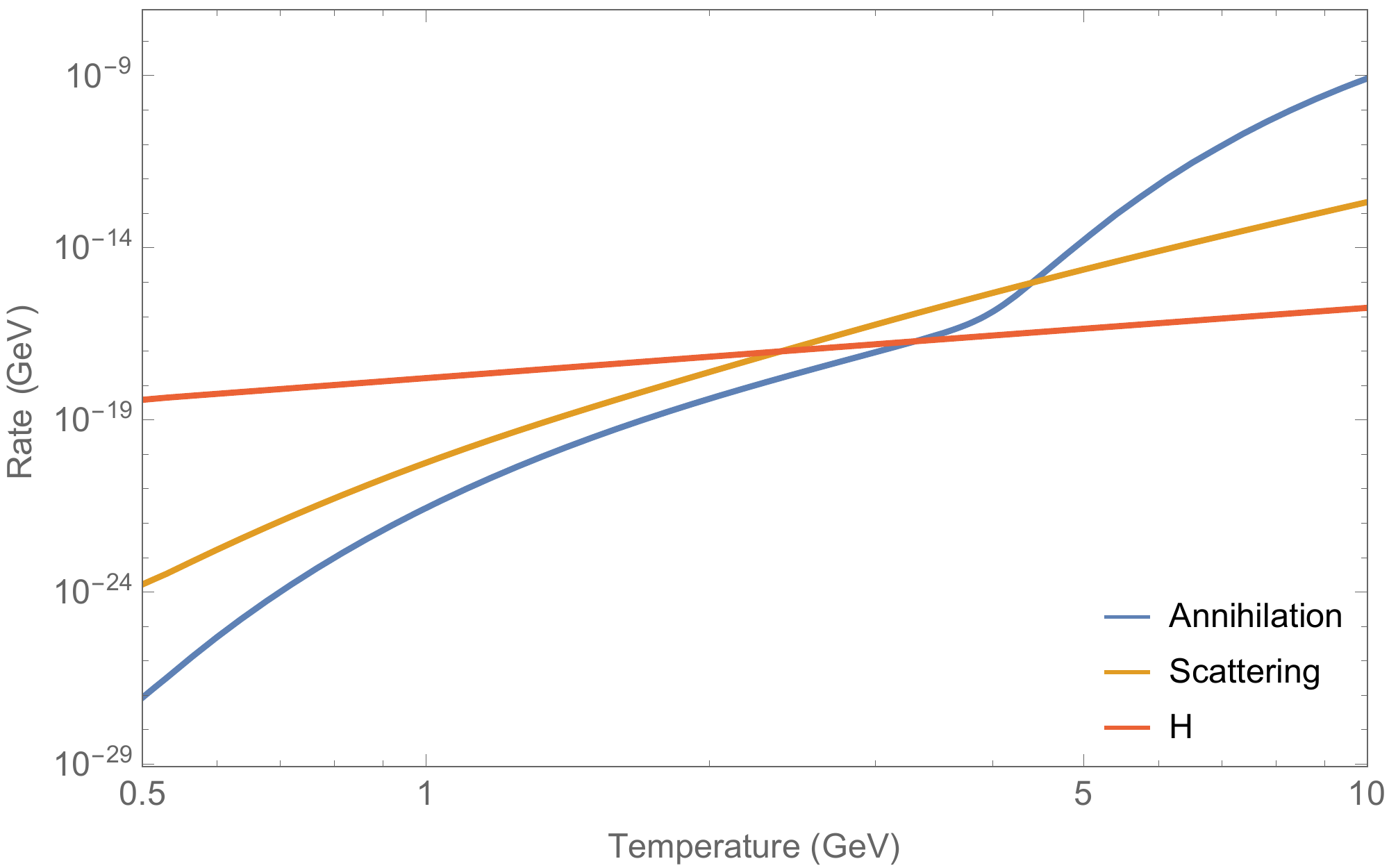}
\centering
\includegraphics[width=.8\textwidth]{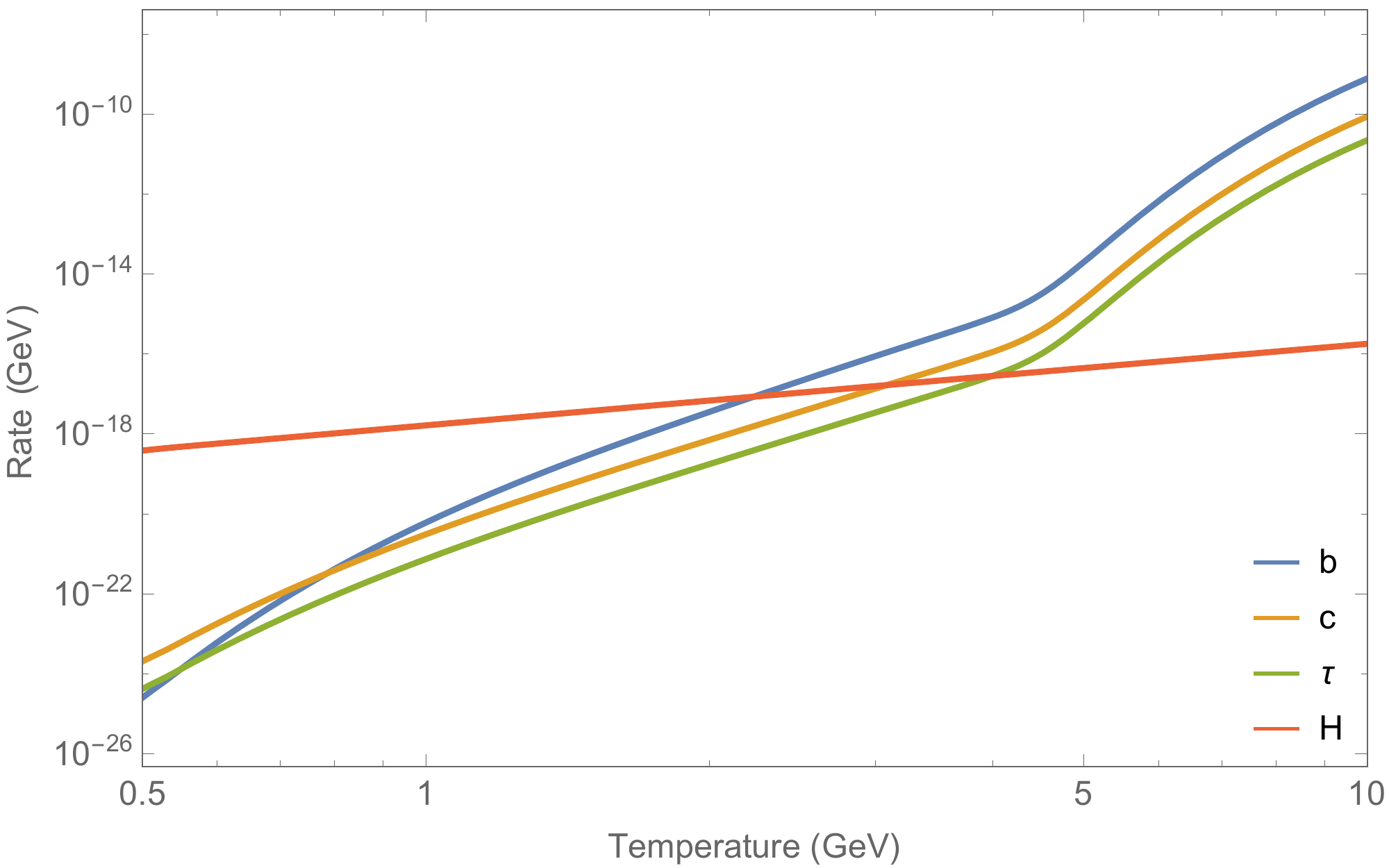}
\caption{Rates of energy density exchange per twin entropy density ($\frac{1}{3s_\text{t}\hat{T}}\frac{dq_{\rm in}}{dt}$) decomposed into contributions from scattering and annihilation (top) and for interactions involving different species of SM fermions (bottom), along with the Hubble parameter, for $f/v=4$. The decoupling temperature is that where the sum of the energy exchange rates equals the Hubble rate, which occurs at $T_{\rm decoup}\approx 2$ GeV.}\label{Fig:decoupling}
\end{figure}






The decoupling temperature depends upon $f/v$, which sets both the mass scale of the twin sector and the strength of the Higgs-mediated coupling. As $f/v$ is increased, decoupling occurs earlier because of the greater Boltzmann suppression, although this is only a relatively small effect that, for $f/v=10$, increases the decoupling temperature by only $4$ GeV. 

When the twin sector is colder than the SM  (which will be important for much of what follows) the heat flow is typically dominated by annihilations of SM into twin particles. However, the energy exchange from elastic scattering can be comparable to that from annihilations, as illustrated in Figure \ref{Fig:decoupling}. Although the energy exchange in an annihilation will generally exceed that of a scattering because all of the energy involved in the process must be transferred, the annihilation rate also becomes more Boltzmann or threshold suppressed when the temperature drops below the mass of the heavier twin particles. It is therefore not always clear that energy transfer through annihilations dominates. 

Decoupling is not exactly instantaneous and there is some range of temperatures over which the rate of heat flow freezes-out. The net heat flow rate $\frac{dq}{dt}$ is greater for larger temperature differences between sectors. The generation of a potentially large temperature difference within this brief epoch of sector decoupling, such as those discussed below in Section \ref{sec:late}, may be cut off when the heat flow rate becomes comparable to the Hubble rate. For a given SM temperature $T$, the minimum twin-sector temperature $\hat{T}_{\text{min}}$ during the decoupling period may be roughly estimated as that which satisfies
\bea
H\sim \frac{1}{3s_\text{t}\hat{T}}\frac{dq}{dt}\Big|_{\hat{T}=\hat{T}_{\text{min}}}. \label{HeatFreeze}
\eea
Twin temperatures colder than $\hat{T}_{\rm min}$ will partially thermalise back to this value. As the participating fermions are not non-relativistic, instantaneous decoupling is not as accurate an approximation as it is, for example, for chemical decoupling of a WIMP, although it is still reliable. 

\begin{figure}[h!]
\centering
\includegraphics[width=.7\textwidth]{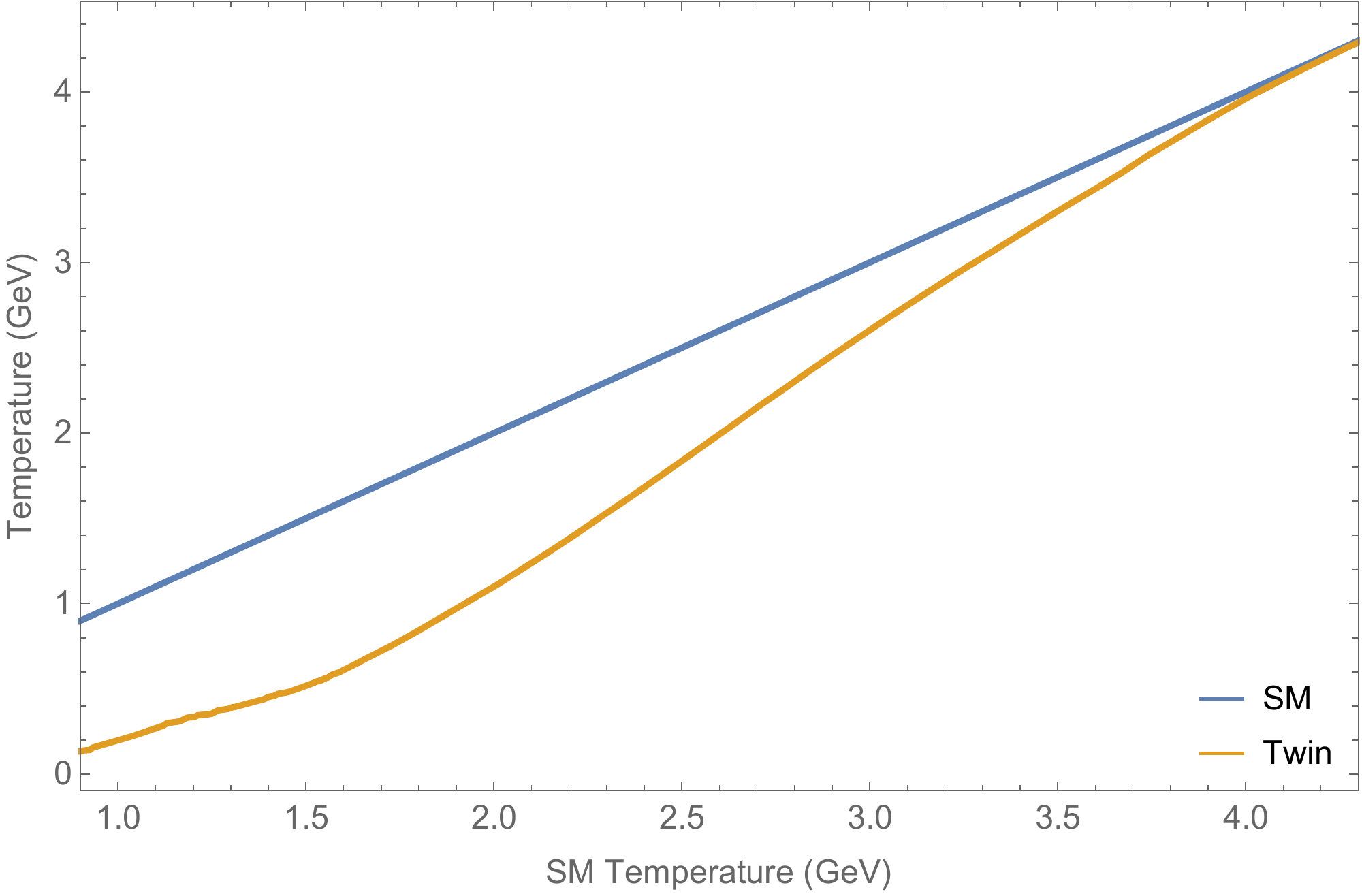}
\caption{Minimum temperature of the twin sector that will not be heated by interactions with a hotter SM plasma, as a function of SM temperature, for $f/v=4$. Also shown is the SM temperature, for reference.}
\label{Fig:decouplingMin}
\end{figure}

In Figure \ref{Fig:decouplingMin}, we show the minimum temperature that the twin sector may have as a function of SM temperature for heat flow to freeze out, estimated using (\ref{HeatFreeze}). Only annihilations have been included in the determination of the minimum temperature, although we have verified that, for these temperatures, the scatterings contribute only $\lesssim 10\%$ to the heat flow. Note that while the energy exchange rate, such as $\frac{1}{\hat{T}}\frac{dq_{\rm in}}{dt}$ in (\ref{ThermEv}), in scattering processes may be faster, the net energy flow rate, or heat flow ($\frac{1}{\hat{T}}\frac{dq}{dt}$ in (\ref{ThermEv})), which is the difference between energy exchange rates into and out of the sector, is actually dominated by annihilations. Generally, we find that decoupling begins at temperatures $\sim 4$ GeV. The temperature difference can reach an order of magnitude without relaxing once the SM temperature drops to $\sim 1$ GeV.  

While the extent of thermal decoupling is temperature dependent, the maximum temperature difference that will not relax grows quickly as the SM temperature drops. Then we may describe the two sectors as being decoupled if, in a given cosmology, all events that raise the temperature of one sector relative to the other (such as the crossing of a mass threshold and the resulting entropy redistribution, the most significant of which is the confinement of colour) induce temperature differences that are too small to partially relax.

At energies $\lesssim 1$ GeV in Figure \ref{Fig:decoupling}, the reliability of the calculation of the heat flow rate diminishes because of the strengthening of the strong coupling and the eventual confinement of colour. Fortunately, for a cooler twin sector, which will be of interest in subsequent sections, annihilations from the SM dominate other processes over most of the parameter space. These are the least sensitive to higher order corrections and non-perturbative effects because of their higher temperature, and hence energy, compared to the potentially cooler twin sector. The range of temperatures illustrated in Figures \ref{Fig:decoupling} and \ref{Fig:decouplingMin} have been selected to roughly illustrate the duration of decoupling, but may extend below the range where the perturbative calculation of the heat flow rate is valid. For example, at temperatures below the twin sector QCDPT, which occurs at $\sim\left(1+\log(\frac{f}{v})\right)$ higher temperatures than in the SM, the partonic calculation of twin quark/anti-quark pair production must be replaced by a hadronic one. Furthermore, the growth of the twin strong coupling necessitates that the quark-Higgs Yukawa couplings be RG evolved to the scale of the energy exchanged, which can induce an $\mathcal{O}(1)$ change to the cross section, although this has only a relatively small effect on the decoupling temperature. It is nevertheless clear that decoupling is mostly complete by then and that these uncertainties are not large enough to affect this conclusion.

In the standard mirror Twin Higgs cosmology, knowing the decoupling temperature tells us how the temperatures of the two sectors will be related at subsequent times. The sectors separately evolve adiabatically after decoupling, though they redshift in the same way and differences in temperature only arise from events that redistribute entropy. Non-minimal cosmological events that could potentially cause the temperatures of each sector to diverge can therefore only be effective if they leave each sector colder than this approximate decoupling temperature.

\subsection{Cosmological Constraints} \label{Sec:Limits}

Given that the twin and Standard Model sectors remain in thermal equilibrium to $\mathcal{O}(\text{GeV})$ temperatures, the simplest mirror Twin Higgs scenario is cosmologically inviable due to the presence of light twin species (photons and neutrinos) with abundances comparable to those of the SM. The cosmological observables through which evidence of light species may be inferred are typically represented by $N_{\rm eff}$, the ``effective number of neutrino species'' in the early universe; their individual masses, which determine their free-streaming distances; and the ``effective mass'' $m_\nu^{\rm eff}$, which parameterises their contribution to the present-day energy density of non-relativistic matter. These observables are probed by both the CMB and large scale structure (LSS).



\subsubsection{Effective number of neutrinos} \label{Sec:Neff}

The parameter $N_{\rm eff}$ describes the amount of radiation-like energy density during the evolution of the CMB anisotropies before photon decoupling. It is defined as the effective number of massless neutrinos with temperature as predicted in the standard cosmology that would give equivalent energy density in radiation:
\begin{equation}
\rho_r=\rho_\gamma+\frac{7}{8}\left(\frac{4}{11}\right)^{4/3}N_{\rm eff}\rho_\gamma,\label{Neffeq}
\end{equation}
where $\rho_r$ is the energy density of radiation and $\rho_\gamma$ is the energy density of photons (the factor of $\Big(\frac{4}{11}\Big)^{4/3}$ arises from the relative reheating of the photons from electron/positron annihilation, which occurs after most of the neutrinos have decoupled, and the factor of $7/8$ is from the opposite spin statistics). A deviation from the Standard Model prediction of $3.046$ \cite{Mangano:2005cc} is denoted by $\Delta N_{\rm eff}=N_{\rm eff}-3.046$. 
This definition of radiation, or equivalently, relativistic degrees of freedom, becomes less clear if the new fields have a non-negligible mass, as we discuss further below. 

We here review the CMB physics of dark radiation, summarising the discussion in \cite{Hou:2011ec}. See also \cite{Ade:2015xua} for further review. The angular size and scale of the first acoustic peak is well-measured and this approximately fixes the scale factor at matter-radiation equality $a_{eq}$. If we imagine fixing all other $\Lambda$CDM parameters, extra radiation would delay the epoch of matter-radiation equality. This would have a pronounced effect on the power spectrum in the vicinity of the first acoustic peak through the early Integrated Sachs-Wolfe (eISW) effect. The modes corresponding to this feature entered the horizon close to matter-radiation equality and the evolution of their potentials is highly sensitive to the radiation energy density. However, the impact of a $\Delta N_{\rm eff}\sim\mathcal{O}(1)$ deviation on the peak height can be simultaneously balanced by increasing the amount of non-relativistic matter, to the extent to which other observations providing independent constraints upon $\Omega_c$ permit (for $\Lambda$CDM$+N_{\rm eff}$, a variation of $\sim 10\%$ in $\Omega_ch^2$ is consistent with present CMB+BAO measurements \cite{Ade:2015xua}, although these variations must be consistent with other observables). 
This degeneracy is not expected to be broken by CMB-S4 \cite{CMB-S4:2016}.

Given that $a_{eq}$ is approximately fixed, the utility of $N_{\rm eff}$ arises because, in simple extensions of the $\Lambda$CDM model, it approximately corresponds to the suppression of power in the small scale CMB anisotropies that arises from Silk damping. The reason for this is roughly that, although the greater expansion rate induced by the extra radiation reduces the time that CMB photons have to diffuse before decoupling, it also reduces the sound horizon size more severely. As the angular size of the sound horizon is determined by the location of the acoustic peaks and is also well measured, the reduction in the sound horizon must be compensated for by a reduction in the angular diameter distance to the CMB. This effectively raises the angular distance over which photon diffusion proceeds and results in a prediction of smoother temperature anisotropies at small scales. This correspondence with the Silk damping allows $N_{\rm eff}$ to be approximately factorised from other parameters and constrained independently, providing a direct observational avenue for detecting the presence of new, massless fields \cite{Hou:2011ec} (see \cite{Brust:2013xpv} for further implications for model building). This relationship arises because the fixing of $a_{eq}$ implies that $N_{\rm eff}$ effectively determines the energy density of the universe, and hence the Hubble rate, during CMB decoupling. Note, however, that further extensions of $\Lambda$CDM may complicate this correspondence, in particular deviations from the standard Big Bang Nucleosynthesis prediction of the primordial helium abundance.

The contribution to $N_{\rm eff}$ (or $\Delta N_{\rm eff}$)
in the mirror Twin Higgs arises from two sources: the twin photons, which can be treated as massless dark radiation with an appropriate twin temperature $T_{\rm eq}^{\rm t}$ at the time of matter-radiation equality, and the twin neutrinos, whose non-zero masses may need to be accounted for. For the twin photons, the contribution to $N_{\rm eff}$ is simple; their equation of state is always $w = 1/3$ and their energy density is given by $g \frac{\pi^2}{30} \left(T^\text{t}_{\rm eq}\right)^4$, where $g = 2$. The twin temperature at matter-radiation equality is found from the SM temperature using comoving entropy conservation,
\begin{equation}
\frac{T^\text{t}_{\rm eq}}{T^\text{SM}_{\rm eq}} = \left(\frac{g^\text{t}_{\star}(T_{\rm decoup})}{g^\text{SM}_{\star}({T}_{\rm decoup})}\right)^{1/3} \left(\frac{g^\text{SM}_{\star}(T^\text{SM}_{\rm eq})}{g^\text{t}_{\star}({T}^\text{t}_{\rm eq})}\right)^{1/3}, \edit{}{fixed labels}
\end{equation}
where the two sectors have the same number of thermalized degrees of freedom by this time. Here, $T^\text{SM}_{\rm eq}$ is the SM photon temperature at matter-radiation equality and ${T}_{\rm decoup}$ is the sector decoupling temperature.

Since neutrinos are massive, their behavior is more complicated. Their equation of state parameter takes on a scale factor dependence which is controlled by their mass. 
In the Standard Model, this sensitivity is negligible because present CMB bounds imply that neutrinos are ultra-relativistic at $a_{\rm eq}$ to good approximation \cite{Ade:2015xua}. However, the factor by which the twin neutrino masses are enhanced may raise them to order $T^t_{\rm eq}$ or greater (see Section \ref{sec:twin} for discussion of the scaling of the masses with $f/v$). 


To better describe the impact of the extra twin (semi-)relativistic degrees of freedom on the CMB, we choose to define $N_{\rm eff}$ through the effects of neutrinos at matter-radiation equality, when the impact on the expansion rate of the universe for most of the period relevant for the evolution of the CMB is greatest. Note that, in their presentation of joint exclusion bounds on $N_{\rm eff}$ and $\sum m_\nu$ (the sum of SM neutrino masses) or $m_\nu^{\rm eff}$ (effective mass contributing to the present-day non-relativistic matter density of an extra sterile neutrino), the Planck collaboration define $N_{\rm eff}$ as the value in (\ref{Neffeq}) at temperatures sufficiently high that the neutrinos are fully relativistic. Our values cannot be directly compared with their analysis, although we consider ours to be a reasonable rough estimate that is more representative of the CMB constraints. The ensuing correction from the finite neutrino masses is, in the cases considered in this work, a small effect anyway.

To determine this correction and provide a definition of $N_{\rm eff}$ that better describes the impact of quasi-relativistic particles on the CMB, we first define the epoch of matter-radiation equality as the time at which the average equation of state parameter of the universe is $\bar{w} = 1/6$ (the equation of state is defined as $\rho=\bar{w} P$, where $\rho$ is energy density and $P$ is pressure). We can express this condition as 
\begin{equation}
\left. \frac{d\ln H}{d\ln a} \right\vert_{a_{eq}}= -\frac{7}{4},
\end{equation}
as in \cite{Dodelson:2005tp}.
Call the quasi-relativistic neutrino energy density $\tilde{\rho}(a)$ with time-evolving equation of state parameter $w(a)$, which is to be balanced against some extra non-relativistic energy density $\Delta \rho_{CDM}(a)\propto a^{-3}$ to keep $a_{eq}$ the same. This amount of non-relativistic energy density $\Delta \rho_{CDM}$ is
\begin{eqnarray}
 \Delta \rho_{CDM} (a_{eq}) = \rho_r(a_{eq}) - \rho_m(a_{eq}) - 2 a_{eq} \left.\frac{d\tilde{\rho}}{da}\right\vert_{a_{eq}} - 7 \tilde{\rho}(a_{eq}),
\end{eqnarray}
where $\rho_r$ and $\rho_m$ are the energy densities of the radiation and non-relativistic matter. 
%
For a perfect fluid, $\frac{d\tilde{\rho}}{da}=-3(1+w(a))\tilde{\rho}/a$ (neglecting the anisotropic stress that is expected only to contribute to a weak phase shift in the CMB \cite{Baumann:2015rya}), 
this results in a Hubble parameter of
\begin{equation}
H^2(a_{eq}) = \frac{2}{3 M_\text{pl}^2} \left[\rho_r(a_{eq}) + 3 w(a_{eq}) \tilde{\rho}(a_{eq})\right].
\end{equation}
This suggests a definition of the effective number of neutrinos, $N_{\rm eff}$, via
\begin{eqnarray}
H^2(a_{eq}) &=& \frac{2}{3 M_\text{pl}^2} \left. \left(\rho_\gamma + N_{\rm eff} \rho_{\nu,m=0}^{th} \right) \right\vert_{a_{eq}} \\
N_{\rm eff} &\equiv& \sum_i\frac{w_i}{1/3} \frac{\rho_i}{\rho_{\nu,m=0}^{th}},
\end{eqnarray}
where $\rho_i$ is the contribution to the energy density from some species $i$ with equation of state parameter $w_i$ and $\rho_{\nu,m=0}^{th}$ is the energy density of a massless neutrino with a thermal distribution in the standard cosmology. Then $3w$ gives the `relativistic fraction' of the energy density. Note that this is simply a ratio of the pressure exerted by the new fields to that of a massless neutrino. The effectiveness of this approximation was discussed in \cite{Archidiacono:2013cha} in the context of thermal axions (while effective at keeping $a_{eq}$ fixed, changes to odd peak heights subsequent to the first are imperfectly cancelled and require further changes to $H_0$ to compensate - see Section \ref{Sec:meff} below).

Calling $T^i_\nu$ the temperature at which the neutrinos in sector $i$ freeze-out and $a^i_\nu$ the corresponding scale factor, then assuming instantaneous decoupling, the phase space number density for scale factor $a$ is given by a redshifted Fermi-Dirac distribution \cite{Jacques:2013xr}
\begin{equation} \label{Eq:nudis}
f^i_\alpha(p) \approx \left[ 1 + e^{pa/\left(a^i_\nu T^i_\nu\right)} \right]^{-1}
\end{equation}
for the $\alpha$ neutrino mass eigenstate in the $i$ sector ($m_\alpha^i\ll T_\nu^i$, so has been dropped). The energy density and pressure are
\begin{eqnarray} 
\rho^i_{\nu_\alpha} &=& \frac{g_\alpha}{2\pi^2} \int_{0}^{\infty} dp \ p^2 \sqrt{p^2 + \left(m^i_\alpha\right)^2}  f^i_\alpha(p) \label{Eq:nuen}\\
P^i_{\nu_\alpha} &=& \frac{g_\alpha}{2\pi^2} \int_{0}^{\infty} dp \ \frac{p^4}{3\sqrt{p^2 + \left(m^i_\alpha\right)^2}}  f^i_\alpha(p) \label{Eq:nupr},
\end{eqnarray}
where $g_\alpha = 2$ is the number of degrees of freedom for a neutrino species. 

Since the neutrino decoupling temperature depends on the strength of the weak interaction as $T_\nu \propto G_F^{-2/3}$, while $G_F \propto v^2$, then the twin neutrino decoupling temperature $T^\text{t}_\nu$ is related to the SM neutrino decoupling temperature $T^\text{SM}_\nu$ by
\bea
T^\text{t}_\nu = (f/v)^{4/3} T^\text{SM}_\nu.\label{nudec}
\eea
We can then simply use (\ref{Eq:nuen}) and (\ref{Eq:nupr}) at matter-radiation equality to find $\Delta N_{\rm eff}$ (assuming instantaneous decoupling). We thus obtain
\begin{eqnarray}
H^2(a_{eq}) &=& \frac{2}{3 M_\text{pl}^2} \left. \left(\rho^\text{SM}_\gamma + 3.046 \rho_{\nu,m=0}^{th} + \rho^\text{t}_\gamma + \sum_\alpha 3 w_{\nu_\alpha} \rho^\text{t}_{\nu_\alpha} \right) \right\vert_{a_{eq}}
\end{eqnarray}
and
\begin{eqnarray} 
\Delta N_{\rm eff} &=& \left(\frac{11}{4}\right)^{4/3} \frac{120}{7 \pi^2 \left(T^\text{SM}\right)^4} \left( \rho^\text{t}_\gamma + \sum_\alpha 3 w^t_{\nu_\alpha} \rho^\text{t}_{\nu_\alpha}\right),\label{CorDelNeff}
\end{eqnarray}
where we now have equation of state parameters $w_{\nu_\alpha}$ for each neutrino, while $\rho^\text{SM}_\gamma$ and $\rho^\text{t}_\gamma$ are the SM and twin photon energy densities, $\rho_{\nu,m=0}^{th}$ and $\rho^\text{t}_{\nu_\alpha}$ are the neutrino energy densities.

\subsubsection{Neutrino masses} \label{Sec:meff}



Because they are so weakly interacting, the neutrinos have a long free-streaming scale given by the distance travelled in a Hubble time $v_\nu/H$, with $v_\nu \propto m_\nu^{-1}$ the speed of the neutrino once it becomes non-relativistic. This defines a free-streaming momentum scale $k_{fs} = \sqrt{\frac{3}{2}} \frac{a H}{v_\nu} \propto m_\nu$, above which neutrinos do not cluster. Below this scale, perturbations in the matter density consist coherently of neutrinos and other matter, but well above it only non-neutrino matter contributes to density perturbations. This results in a suppression of the matter power spectrum on large scales which is proportional to the fraction of energy density in the free-streaming matter. Since this occurs at late times when neutrinos are non-relativistic, the energy density is simply $\rho_{\nu_\alpha} = n_{\nu_\alpha} m_{\nu_\alpha}$ for each neutrino species $\alpha$, where $n_{\nu_\alpha}$ is the number density. 
Constraints on the sum of neutrino masses then come from the observations of power on small scales, which is suppressed relative to that expected for massless neutrinos by a factor $\appropto 1-8 f_\nu$, where $f_\nu = \Omega_\nu/\Omega_m$ is the fraction of non-relativistic energy in neutrinos at late times \cite{Lesgourgues:2012uu}.


More generally, inferences of the matter power spectrum constrain the present-day energy density fraction of free-streaming species that do not cluster on small scales and have since become non-relativisitic, $\Omega_\nu=(\sum m_\nu + m_\nu^{\rm eff})/(94.1\,\text{eV})$, where $\sum m_\nu$ is the sum of SM neutrino masses and $m_\nu^{\rm eff}$ is the sum of twin neutrino masses weighted by their number density
\begin{equation}
m_\nu^{\rm eff} = \frac{n^\text{t}_\nu}{n^\text{SM}_\nu} \sum_\alpha m^\text{t}_{\nu_\alpha}.
\end{equation} 
Here $n^{\text{t}}_\nu$ is the number density of a relic twin neutrino flavour and $n^\text{{SM}}_\nu$ is that for a SM neutrino. It is assumed that the neutrinos have been thermally produced as hot relics.

The relic abundance of a neutrino species is given by its number density when it decoupled, diluted by the factor by which the universe has since expanded. The scale factors at which neutrino decoupling occurs in the two sectors, $a^\text{SM}_\nu$ and $a^\text{t}_\nu$ can be determined from (\ref{nudec}), the relative temperatures in the two sectors and comoving entropy conservation, to obtain 
\begin{eqnarray}
a^\text{t}_\nu &=& a^\text{SM}_\nu \left(\frac{v}{f}\right)^{4/3} 
\left( \frac{g^\text{t}_\star\left(T_{\rm decoup}\right)}{g^\text{SM}_\star\left(T_{\rm decoup}\right)}\right)^{1/3}
\end{eqnarray}
where the same mass thresholds have been assumed in each sector below their neutrino decoupling temperatures, so that $g^\text{SM}_\star\left(T^\text{SM}_\nu\right)=g^\text{t}_\star\left(T^\text{t}_\nu\right)$. The neutrino number densities are then
\begin{eqnarray}
\frac{n^\text{t}_\nu}{n_\nu^\text{SM}} =\left(\frac{T^\text{t}_\nu a^\text{t}_\nu}{T^\text{SM}_\nu a^\text{SM}_\nu}\right)^3 
=\frac{g^\text{t}_\star\left(T_{\rm decoup}\right)}{g^\text{SM}_\star\left(T_{\rm decoup}\right)}.
\end{eqnarray}
For $f/v$ from $3$ to $10$ and using $T_{\rm decoup} \sim 2 - 6$ GeV from Section \ref{Sec:Decoupling}, we find \\ $g^\text{t}_\star\left(T_{\rm decoup}\right)/\ g^\text{SM}_\star\left(T_{\rm decoup}\right) \sim 0.8$ and thus arrive at 
\begin{equation}
m_\nu^{\rm eff} \approx 0.8 \left(\frac{f}{v}\right)^n \sum_\alpha m^\text{SM}_{\nu_\alpha},
\end{equation}
where $n = 1$ for Dirac masses and $n = 2$ for Majorana masses. 

If they are sufficiently light and hot, the twin neutrinos only affect the CMB as dark radiation and their masses may then only be inferred from tests of the matter power spectrum. However, if heavier and colder, they are better described as a hot dark matter component. Their impact on the CMB is discussed in \cite{Dodelson:1995es}, where the shape of the power spectrum can depend upon the individual neutrino kinetic energies through their characteristic free-streaming lengths. The early Integrated Sachs-Wolfe effect (eISW) is also sensitive to the masses if the neutrinos become non-relativistic during decoupling (thereby affecting the radiation energy density and the growth of inhomogeneities) \cite{Lesgourgues:2012uu}.

There is a significant degeneracy in cosmological fits to the CMB between $\Omega_m$ and $H_0$ (the Hubble constant) \cite{2012JCAP...04..027H}, where raising the non-relativistic matter fraction, such as with nonrelativistic neutrinos, can be accommodated by a decrease in $H_0$ (or equivalently, the dark energy density), which keeps the angular diameter distance to the CMB approximately fixed. This degeneracy can be broken by measurements of the baryon acoustic oscillations (BAOs), which are sensitive to the expansion rate of the late universe and provide an independent measurement of $\Omega_m$ and $H_0$. It is through combination with these results that bounds from Planck on neutrino masses are strongest \cite{Ade:2015xua}.

\subsubsection{Bounds}\label{Bounds}

The authors are unaware of any specialised analysis of the present and projected future cosmological constraints on scenarios with both massless dark radiation and additional light, semi-relativistic sterile neutrinos. In the absence of this, we use bounds from \cite{Ade:2015xua} as a rough indication of the present level of sensitivity to these parameters, which we nevertheless expect to be a reliable indication of the (in)viability of this model. The 95\% confidence limits on these parameters are $N_{\rm eff} = 3.2 \pm 0.5$ and $\sum m_\nu < 0.32 \text{ eV}$ when each are constrained separately with the other fixed. This, of course, overlooks correlations between the impacts of masses and $\Delta N_{\text{eff}}$ on the CMB and LSS. Bounds on an additional sterile neutrino as the only source of dark radiation are also presented with number density, or equivalently, contribution to $\Delta N_{\text{eff}}$, left to float. These are similar to the limit on $\sum m_\nu$. It was found in \cite{DiValentino:2016ikp} that, allowing $\sum m_\nu$ and $m_\nu^{\text{eff}}$ to float independently for a single extra sterile neutrino, the bound mildly relaxes to $m_\nu^{\rm eff}\lesssim 1$ eV, although the bound may be stronger depending on the combination of data sets chosen (the lensing power spectrum presently prefers higher neutrino masses and raises the combined bounds if included). Other bounds from LSS on $\sum m_\nu$ exist and are potentially stronger than those placed from the CMB, possibly as low as $m_\nu^{\rm eff}\lesssim 0.05$ eV, again depending on data sets combined (see \cite{Costanzi:2014tna}, \cite{Gariazzo:2015rra}), although these are subject to greater uncertainties in the inference of the power spectra of dark matter halos from galaxies surveys and the Ly$\alpha$ forest.

It must also be noted that the shape of the CMB temperature anisotropies depends upon both the mass of individual neutrino components (through their free-streaming distance) and their contribution to the energy density of the nonrelativistic matter that does not cluster on small scales. However, it is not expected that improvements in bounds on the former will be made from improved measurements of the primary CMB itself, but rather from weak lensing of the CMB, in conjunction with future measurements from DESI of the BAOs to break degeneracy with $\Omega_m$. The lensing spectrum, like inferences of the matter power spectrum made in galaxy surveys, is expected to measure the suppression of small scale power and therefore to strengthen constraints upon $m_\nu^{\rm eff}$, rather than the individual neutrino masses. 
One of the goals of CMB-S4 will be the detection of neutrino masses, given the present lower bound $\sum m_\nu\gtrsim 0.06$ eV from oscillations. Projected bounds are as low as $\sim 0.02$ eV \cite{CMB-S4:2016}, although this assumes no extra dark radiation or sterile neutrinos. A projection of the joint bound on $N_{\rm eff}$ (from extra massless dark radiation) and $m_\nu^{\rm eff}$ combining improved measurements CMB temperature measurements, lensing and BAOs indicates a limit of $m_\nu^{\rm eff}\lesssim 0.1$ eV at $1\sigma$ \cite{CMB-S4:2016}. Any contribution from additional states to $m_\nu^{\rm eff}$ may therefore be testable and bounded by the excess of the neutrino mass inference over the minimum neutrino mass, although laboratory measurements or measurements of $\Delta N_{\rm eff}$ will be required to further ascertain the contribution from the new particles. 

Constraints on $\Delta N_{\rm eff}$ from improved measurements of the damping tail as part of CMB-S4 are projected to be $\sim 0.02-0.05$ at $1\sigma$ \cite{CMB-S4:2016}. 
In the following sections, we use an optimistic estimate of $0.02$ for its reach in order to identify as much of the potentially testable parameter space as possible.

\begin{figure}[h!]
\centering
\includegraphics[width=1.0\linewidth]{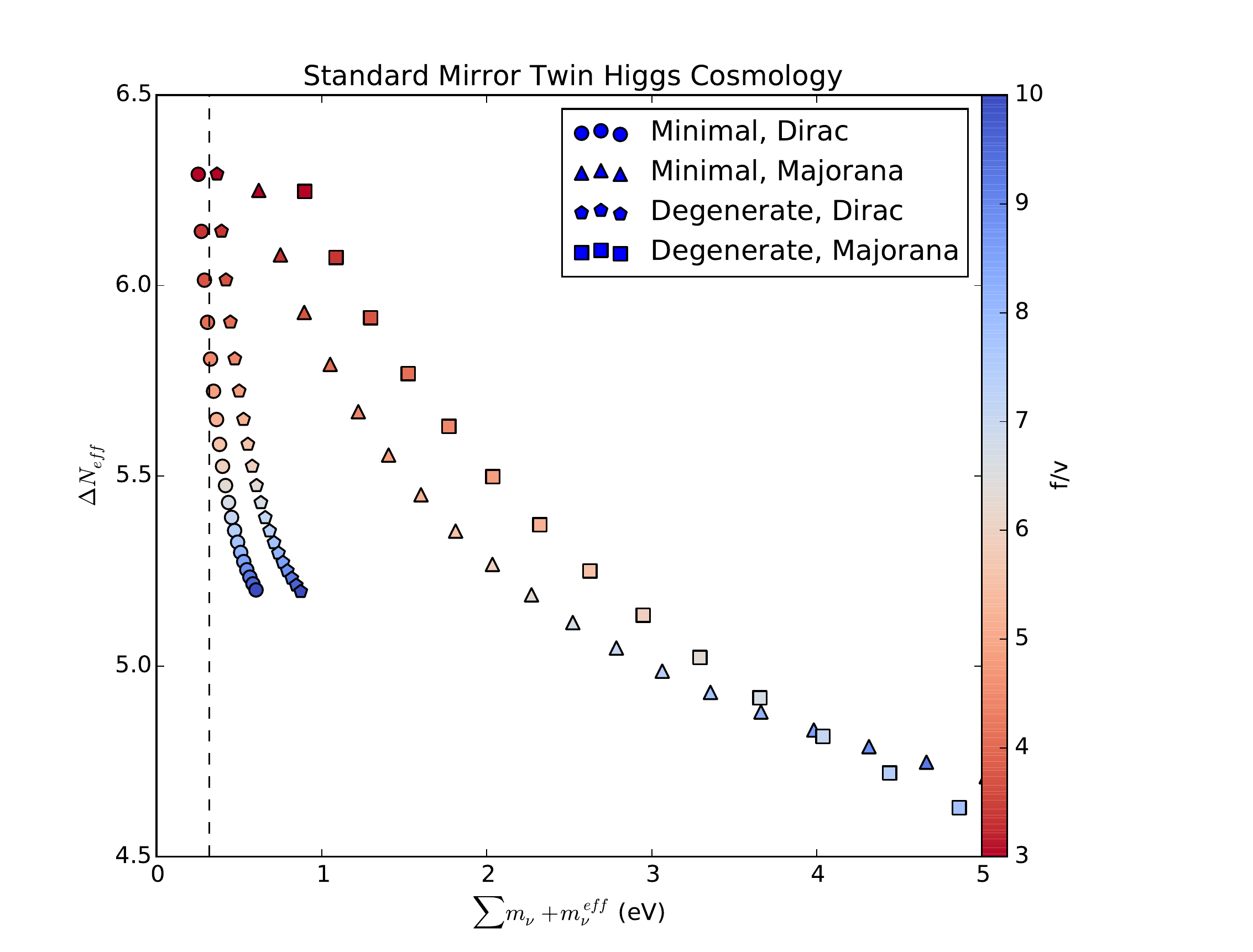}
\caption{Predicted values of $\Delta N_{\rm eff}$ and $\sum m_\nu + m_\nu^{\rm eff}$ for minimal and degenerate neutrino mass spectra with both Dirac and Majorana masses for $f/v$ from 3 to 10. The Planck 2015 constraint\cite{Ade:2015xua} is the dashed line; the corresponding $N_{\rm eff}$ upper bound is well below the bottom of the plot. All points are excluded by the combination of bounds on $\Delta N_{\rm eff}$ and $\sum m_\nu + m_\nu^{\rm eff}$.}
\label{Fig:mirrorPred}
\end{figure} 

To estimate the impact of current and projected CMB limits on the mirror Twin Higgs, we consider two scenarios: the minimal Standard Model neutrino mass spectrum of $m_\nu = \left[0.0,0.009 \text{ eV},0.06 \text{ eV}\right]$ and a degenerate spectrum of $m_\nu = \left[0.1\text{ eV} ,0.1\text{ eV},0.1 \text{ eV}\right]/3$ from \cite{Ade:2015xua}. In Figure \ref{Fig:mirrorPred} we plot the predictions of the mirror Twin Higgs for $\Delta N_{\rm eff}$ and $m_\nu^{\rm eff}$ for both types of spectra, as well as for both Dirac and Majorana masses (which scale differently with $f/v$). As is plainly evident, the mirror Twin Higgs is ruled out cosmologically, no matter the choices of neutrino masses one makes, if only for the presence of the twin photon.
In the standard cosmology, the twin sector will have roughly the same temperature as the SM, giving $4.6\lesssim\Delta N_\text{eff}\lesssim 6.3$ for $f/v<10$, according to the definition of (\ref{CorDelNeff}). This range depends upon $f/v$ through the twin neutrino decoupling temperature (\ref{nudec}), which determines the extent to which the twin photons are reheated relative to the twin neutrinos after twin electron/positron annihilations. This is sufficiently large that even the cold dark matter fraction cannot be adjusted to keep matter-radiation equality fixed, resulting inevitably in changes to the height and shape of the first acoustic peak. The energy density in neutrinos is predicted to be above the present observational upper bounds for most neutrino mass configurations, with the exception of the minimal values permitted by neutrino oscillation measurements with $f/v\lesssim 6$. We therefore discuss cosmological mechanisms in which the twin radiation is diluted to levels compatible with these observational bounds in the subsequent sections of this paper. 

\section{Reheating by the decay of a scalar field} \label{sec:late}

We now turn to simple scenarios that reconcile the mirror Twin Higgs with cosmological bounds, while taking care to respect the softly-broken $\mathbb{Z}_2$ symmetry. We begin with the out-of-equilibrium decay of a particle with symmetric couplings to the Standard Model and twin sectors, in which the desired asymmetry is generated kinematically. That is to say, the dimensionless couplings between the decaying particle and the two sectors are equal, and asymmetric energy deposition into the two sectors is a direct consequence of the asymmetric mass scales. In this respect, the scenario is philosophically similar to $N$naturalness \cite{Arkani-Hamed:2016rle}, albeit with a parsimonious $N=2$ sectors. See also \cite{Reece:2015lch}, \cite{Randall:2015xza} and \cite {Adshead:2016xxj} for other recent related ideas of using long-lived particles for the dilution of dark sectors.

For simplicity, here we will focus on the case of a real scalar $X$ coupled symmetrically to the $A$ and $B$ sector Higgs doublets. Due to the difference in masses between the sectors after electroweak symmetry breaking, simple kinematic effects give $X$ a larger branching ratio into the Standard Model. This occurs over a range of $X$ masses within a few decades of the weak scale. If $X$ decays out-of-equilibrium below the decoupling temperature of the two sectors, this injects different amounts of energy into the two sectors, effectively suppressing the temperature of the twin sector relative to the Standard Model. This relative cooling suppresses the contribution of the light degrees of freedom of the mirror Twin Higgs to below cosmological bounds. Insofar as the asymmetry is driven entirely by kinematic effects arising from $v \ll f$, the resulting temperature inequality between the two sectors is proportional to powers of $v/f$.

The requisite suppression of the twin sector temperature relative to the Standard Model temperature necessitates that the $X$ dominate the cosmology before it decays. Our main discussion will follow the simplest case of an $X$ which dominates absolutely before it decays, comprising all of the energy density of the universe and effectively acting as a `reheaton'. Afterwards, we will discuss the possibility of a `thermal history' for $X$ -- a scenario where $X$ is in thermal equilibrium with the two sectors, then chemically decouples at some high temperature and grows to dominate the cosmology before it decays. This scheme will result in additional stringent constraints on the viable parameter space.

\subsection{Asymmetric Reheating} \label{Sec:reheaton}

A $\mathbb{Z}_2$-even scalar $X$ which is a total singlet under the SM and twin gauge groups admits the renormalisable interactions 
\begin{equation} \label{eq:portal}
V \supset \lambda_x X (X + x) \left(\left|H_A\right|^2 + \left|H_B\right|^2\right)+\frac{1}{2}m_X^2X^2,
\end{equation}
where $m_X$ is the mass of $X$ (neglecting corrections from mixing that will be shown below to be tiny), $\lambda_x$ is a dimensionless coupling and $x$ is a dimensionful parameter, which one may imagine identifying as a vacuum expectation value (vev) of $X$ in an UV theory. Note that these interactions preserve the accidental $SU(4)$ symmetry of the Twin Higgs. The $X$ field may additionally possess self-interactions, which we omit here as they do not play a significant role in what follows.

The interactions in (\ref{eq:portal}) allow $X$ to decay into light states in the Standard Model and twin sectors. If $X$ reheats the universe through out-of-equilibrium decays, the reheating temperatures of the two sectors will be determined by its partial decay widths, assuming that the decay products do not equilibrate. In the instantaneous decay approximation, $X$ decays when the Hubble parameter falls to its decay rate $\Gamma_X \sim H$. As we will show in Section \ref{Sec:CMBfx}, in order to evade cosmological constraints we need the $X$ to decay mostly into the SM, so we may estimate $\Gamma_X \sim \Gamma(X \rightarrow \text{SM})$. Then the energy that was contained in the $X$ is transferred into radiation energy density, with the resulting temperature of the radiation given by (see \cite{Kolb:1990vq})
%
\begin{eqnarray}
T \sim 1.2\sqrt{\frac{\Gamma_X \mpl}{\sqrt{g_\star}}}\label{ReheatTemp}
\end{eqnarray}
where $g_\star$ is the effective number of relativistic degrees of freedom, as defined in Section \ref{sec:thermal}, of the particles that are being reheated. Our numerical calculation of the reheating temperature, which will be presented in Section \ref{Sec:cosmosim}, indicates that the approximation $T \sim 0.1\sqrt{\Gamma_X \mpl}$ reliably reproduces the reheating temperature over the range of interest. 

As shown in Section \ref{Sec:Decoupling}, the two sectors thermally decouple when the temperature falls below $T_{\rm decoup} \sim 1 \text{ GeV}$, so reheating must take place to below this temperature. At even lower temperatures, big bang nucleosynthesis (BBN) places strong constraints on energy injected into the SM at temperatures below $\mathcal{O}(1-10)$ MeV \cite{deSalas:2015glj}.
Requiring that the SM reheating temperature is above $\sim 10$ MeV, these constraints on the SM reheating temperature become constraints on the decay rate of the $X$ into the SM, which in the above approximation becomes
\begin{equation}
5\times 10^{-21} \ \text{GeV} \lesssim \Gamma_X \lesssim 3\times 10^{-16} \ \text{GeV}.\label{ReheatRange}
\end{equation}
This then constrains the couplings $\lambda_x$ and $x$ of the $X$ to the Higgs sector. Importantly, it means that $X$ must couple very weakly, in order to be long-lived enough to reheat to a low temperature, as will be shown below.


The asymmetry in partial widths arises from different effects depending upon the mass of $X$. For masses below the SM Higgs threshold, it is predominantly differences in mass mixing with the two Higgs doublets that produces the asymmetry, where the size of the mixing angles determines the effective coupling of $X$ to the SM and twin particles and therefore its branching fractions. For masses below the twin scale, the relative size of the mixing scales inversely with the vevs in each sector. Thus the hierarchy $v \ll f$ already present in the Higgs sector can automatically gives rise to a hierarchy in partial widths. Note that additional threshold effects can enhance the asymmetry further, in particular when $X$ has mass above threshold for a significant decay channel in the SM, but below the corresponding mass threshold in the twin sector. Decays into on-shell Higgses complicate this picture further.
In what follows, we first give an analytic calculation of the mass mixing effect, then present a more precise calculation of the decay widths into each sector.


To lowest order, $X$ decays via its interactions with the SM and twin Higgs, and only to other fermions and gauge bosons through its mass mixing with the Higgs scalars. Expanding the $X$ potential after the $SU(4)$ is spontaneously broken, the mixing term between $X$ and $h_A$ in the scalar mass matrix is $\sqrt{2}\lambda_x x v_A$, while that between $X$ and $h_B$ is $\sqrt{2}\lambda_x x v_B$. The $h_A$ and $h_B$ components of the $X$ mass eigenstate, which we denote respectively as $\delta_{XA}$ and $\delta_{XB}$, can then be determined. The expressions for the mixing angles are in general complicated, but they simplify in limits $m_X < f$ and $m_X \gg f$:
\begin{equation} \label{Eq:mixing} 
\left(\delta_{XA}, \delta_{XB}\right) \approx 
\begin{cases}
\frac{4 \lambda_x x v_A}{m_X^2 - m_h^2} \left(\frac{1}{\sqrt{2}},\frac{v_A}{f}\right) & m_X < f \\
\frac{\lambda_x x f}{m_X^2} \left(\frac{\sqrt{2} v_A}{f},1\right) & m_X \gg f
\end{cases}
\end{equation}
to lowest order in $(v/f)^2$ and $\kappa/\lambda$. The partial width for the decay of $X$ into SM states (excluding the Higgs) is
\begin{equation}
\Gamma(X \rightarrow \text{SM}) \approx \left|\delta_{XA}\right|^2 \Gamma_h(m_h = m_X),
\end{equation}
where $\Gamma_h(m_h = m_X)$ denotes the decay width of a SM Higgs if it were to have mass $m_X$. Note that the Higgs partial width must be computed using the vev $v_A \approx v/\sqrt{2}$ to determine the masses and couplings of the SM particles. The partial width of the $X$ into twin states is computed the same way using $\delta_{XB}$ and the vev $v_B \approx f/\sqrt{2}$.

From the mixing angles (\ref{Eq:mixing}), it is already apparent over what mass range asymmetric reheating from $X$ decays will work. These give
\begin{equation} 
\frac{\Gamma(X \rightarrow \text{SM})}{\Gamma(X \rightarrow \text{Twin})} \sim
\begin{cases}
f^2/v_A^2 \gg 1 & m_X < f \\
v_A^2/f^2 \ll 1 & m_X \gg f.
\end{cases}
\end{equation}
Thus when the mass of $X$ is less than the twin scale, the Standard Model will be reheated to a higher temperature than the twin sector, but in the large mass limit this mechanism works in the opposite direction and would appear to lead to preferential reheating of the twin sector. 

More precise statements about the relative branching ratios and resulting temperatures require additional care. In addition to decaying through mass mixing, $X$ can decay into the Higgs mass eigenstates themselves if above threshold. 
As the energy is ultimately transferred to the SM and twin sectors, we then need to consider how these states decay and account for the further mixing of the Higgs mass eigenstates into Higgs gauge eigenstates.

For $m_X > 2 m_h$, decay can occur into the lighter (SM-like) Higgs mass eigenstate $h$ with partial width
\begin{equation}
\Gamma(X\rightarrow hh)\approx\frac{\lambda_x^2 x^2}{16\pi m_X}\sqrt{1-\left(\frac{2m_h}{m_X}\right)^2}\label{Eq:Xtohh}.
\end{equation}
Similarly, for $m_X>2m_H$, decays can proceed into $HH$ with a similar partial width, but with the $h$ mass replaced with that of the $H$.
Above the intermediate threshold $m_X > m_h + m_H$, there is also the mixed decay
\begin{equation}
\Gamma(X\rightarrow hH)\approx\frac{\lambda_x^2}{2\pi m_X}\sqrt{1-\left(\frac{m_H+m_h}{m_X}\right)^2}(f\delta_{AX}+2v_A\delta_{BX})^2. 
\end{equation}
Here, $\delta_{AX} \approx-\delta_{hA}\delta_{XA}-\delta_{hB}\delta_{XB}$ is the component of the $h_A$ gauge eigenstate in the $X$ mass eigenstate and $\delta_{BX}\approx\delta_{hB}\delta_{XA}-\delta_{hA}\delta_{XB}$ is the corresponding component of the $h_B$ gauge eigenstate, where $\delta_{hA}$ and $\delta_{hB}$ are, respectively, the components of the SM Higgs in the $h_A$ and $h_B$ gauge eigenstates to zeroth order in $\lambda_x$. Combining all ingredients, this decay width is of order $\lambda_x^4x^2$. Since it is only the total decay width that is constrained to be small by the demand that the SM reheating temperature lie in the required window, this fixes only a product of $\lambda_x$ and $x$. If $x\sim v$, then the mixed decay to $hH$ is effectively second order in the small coupling $\lambda_x^2$ and can be neglected relative to the other partial widths. Conversely if $x\ll v$, then $\lambda_x$ is much larger and this decay cannot be neglected. In what follows we will work in the region of parameter space where mixed decays to $hH$ are negligible.

The rate of heat flow into each sector may be well approximated by adding the decay rates of $X$ into each channel and weighting these by the fraction of energy transferred into the particular sector. Of course, when $X$ decays into Higgs particles, these in turn decay out of equilibrium into both the Standard Model and twin sectors. As the Higgs decays are almost instantaneous, the fraction of energy transferred into each sector is simply that carried by the Higgs decay products multiplied by their branching fractions for each sector. 
The total rate at which $X$ particles are transferred into the SM plasma is 
\begin{multline}
W(X\rightarrow SM)\approx \Gamma(X\rightarrow SM)+\Gamma(X\rightarrow hh)Br(h\rightarrow SM)\\ \hspace{0cm}\hbox{}+  \Gamma(X\rightarrow HH)(Br(H\rightarrow SM)+Br(H\rightarrow hh)Br(h\rightarrow SM)).\label{Xrate}
\end{multline}
The corresponding rate for energy deposition into the twin sectors is simply given by the replacement of ${ SM} \mapsto {\text{Twin}}$. The first term is the rate at which $X$ decays directly into the SM through mass mixing with the Higgs. The second is the fraction of $X$ energy that is transferred into lighter Higgs states that subsequently decay into the SM. The third is the analogous term for decays into the heavy Higgs, where cascade decays of the $H$ into the $h$ and subsequently other SM particles must be included. Note that decays of the heavy Higgs into the light Higgs make up a majority of decay width, because of the large quartic coupling required for the twin Higgs potential. 

Below the $hh$ threshold, it is possible for $X$ to decay via one on-shell and one off-shell Higgs boson. The partial width for off-shell Higgs production was calculated for $X \rightarrow hh^* \rightarrow hb\bar{b}$ and found to be negligible compared to two-body decays through mass mixing and so we omit three-body decay widths in what follows.

\begin{figure}[t]
\centering
\includegraphics[width=0.8\linewidth]{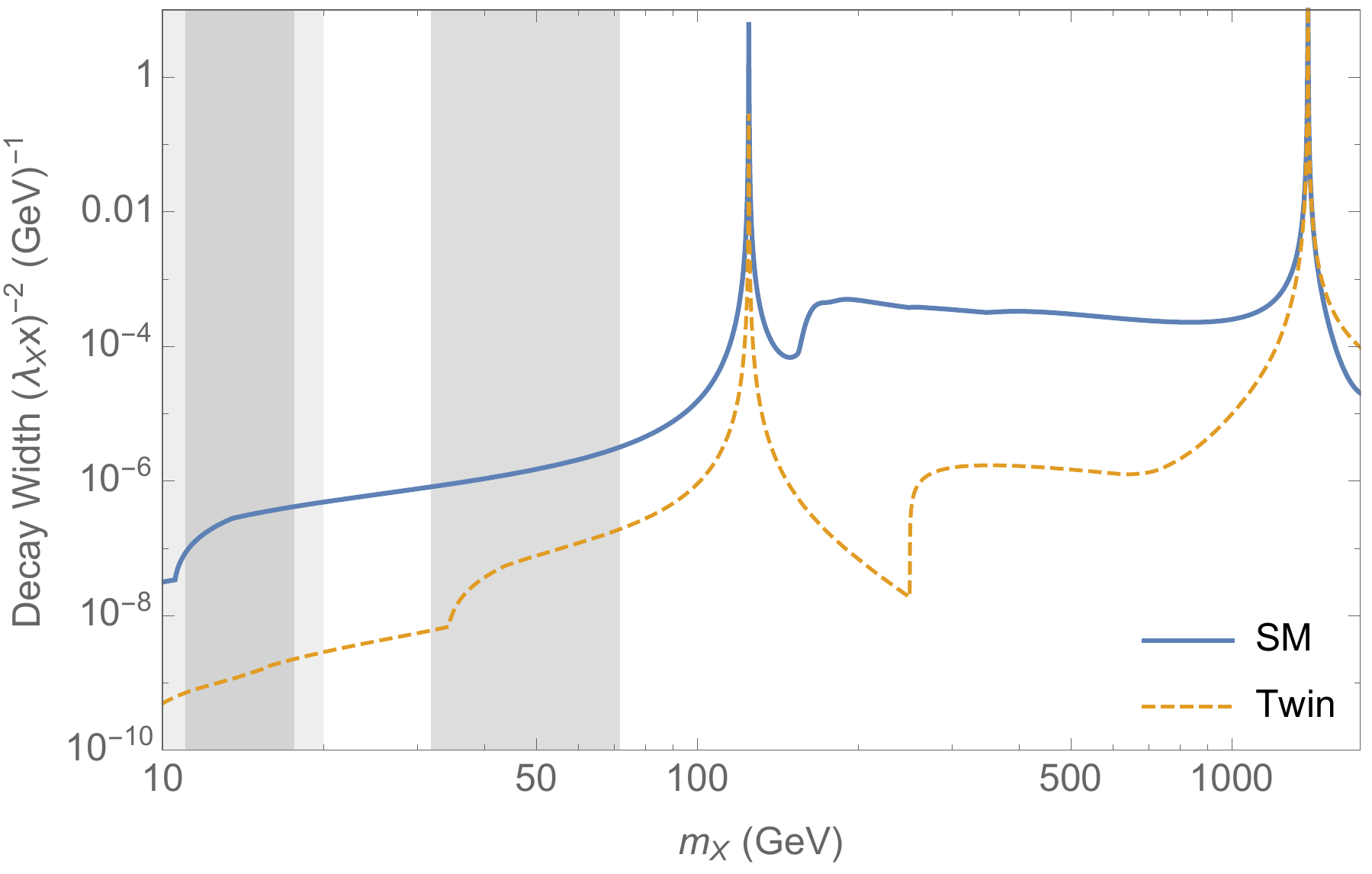}
\caption{The partial widths of the $X$ into the SM (solid blue line) and twin sector (dashed orange) for $f/v = 3$ in units of $(\lambda_x x)^2$. The light gray bands indicate regions of QCD-related uncertainty in the SM calculation, while the darker gray bands indicate the corresponding regions of uncertainty for the twin calculation.}
\label{Fig:decayWidths}
\end{figure}
\begin{figure}[t]
\centering
\includegraphics[width=0.7\linewidth]{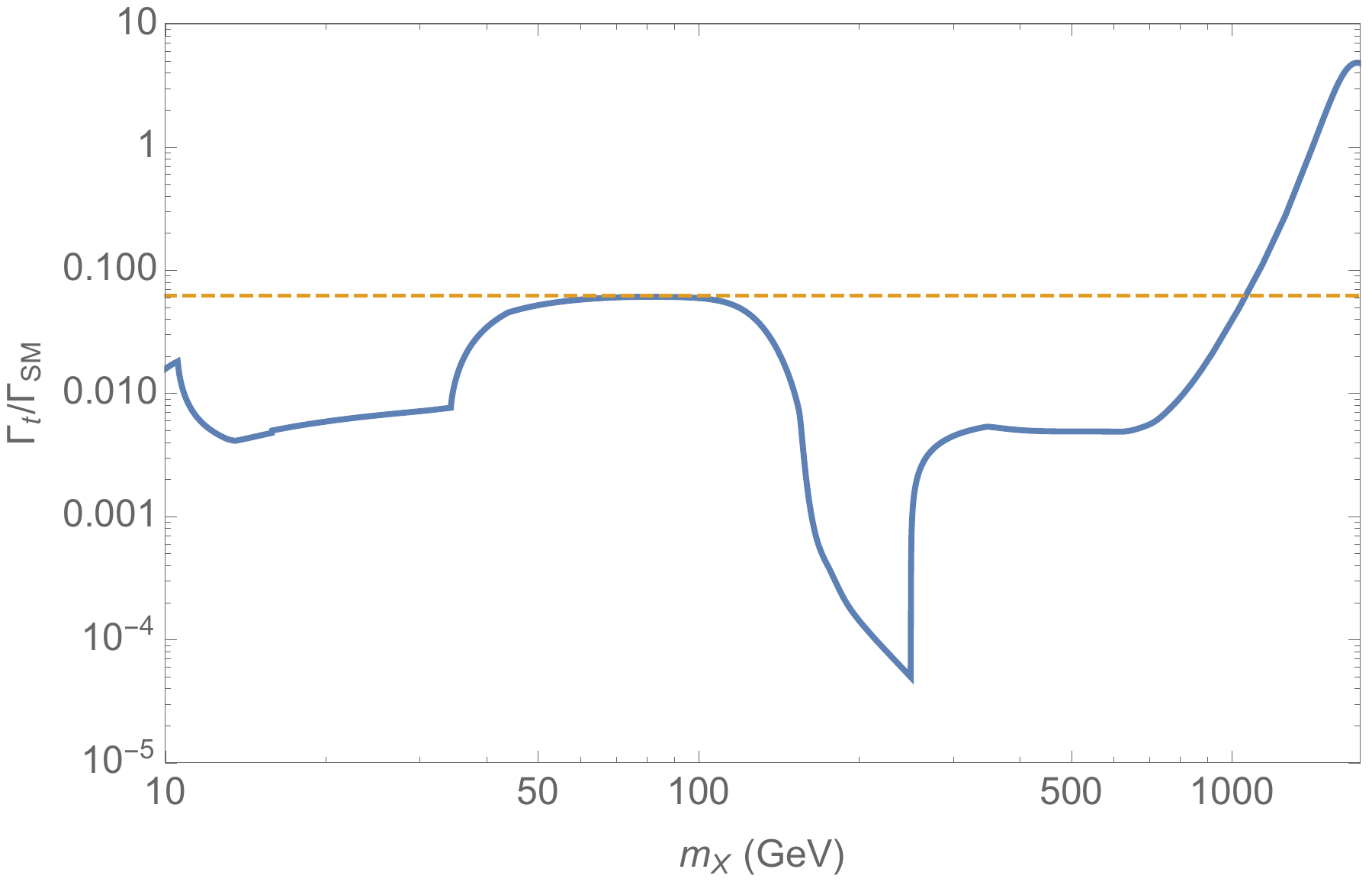}
\caption{The ratio of branching fractions of the $X$ into the SM and twin sectors at $f/v=3$. The dashed line gives the expected $\left(v/f\right)^2$ scaling from the mass mixing; deviations are due to various mass threshold effects.}
\label{Fig:brRatio}
\end{figure}

Ultimately, the complete partial widths for the decay of $X$ into the Standard Model and twin sectors includes the sum of decays into Higgs bosons $h$ and $H$ and direct decays into the fermions and gauge bosons of the two sectors. We compute the latter to an intended level of accuracy of $\sim 10\%$ (including, e.g., NLO QCD corrections to decays into light-flavor quarks), mostly following \cite{Djouadi:2005gi}. The resulting partial widths into the Standard Model and twin sectors are shown as a function of $m_X$ in Figure \ref{Fig:decayWidths} with the ratio of branching fractions displayed in Figure \ref{Fig:brRatio}.  

Over much of the space below the Higgs mass, the branching ratio exhibits the expected $(f/v)^2$ scaling from the mass mixing. Below $\sim 40$ GeV, suppression of the twin partial width arises because the twin bottom quark pair production threshold is crossed. As $m_X$ nears $m_h$, the SM branching fraction grows by $\sim 4$ orders of magnitude as the $WW^*$, $ZZ^*$, and then $WW$ and $ZZ$ decays go above threshold. Since the analogous thresholds are at much higher energies in the twin sector, the enhancement is not paralleled by decays into the twin sector until $m_X$ is close to the twin scale. There is therefore a large range of masses $m_h \lesssim m_X \lesssim m_H$ over which the SM branching fraction dominates by several orders of magnitude.

Above the $X\rightarrow hh$ threshold, the ratio of decay widths is roughly constant in mass up to the $HH$ threshold. The twin sector decay rate is dominated by decays of on-shell light Higgs into twin states, $\Gamma(X \rightarrow \text{Twin}) \approx \Gamma(X\rightarrow hh)\text{Br}(h\rightarrow\text{Twin}) \propto 1/m_X $ as in (\ref{Eq:Xtohh}). If the SM were also predominantly reheated through this channel, then the ratio of branching fractions would again be approximately $\delta_{hA}^2/\delta_{hB}^2\approx (f/v)^2$. However, the SM decay width also receives a larger contribution from decays through mass mixing between the $X$ and the Higgs gauge eigenstates. 

For masses $m_X>2m_h$, decays through mass mixing are dominated by the SM $WW$ and $ZZ$ channels. In this mass region, the decay rate of a Higgs into longitudinally polarized vector bosons scales as $\Gamma(h\rightarrow WW,ZZ)\sim m_X^3$, but the mixing angle scales as $\delta_{AX}^2\sim 1/m_X^4$ (as in (\ref{Eq:mixing})), resulting in the same $\sim 1/m_X$ scaling and thus a roughly constant ratio in this range of masses. 
Near $m_X \sim 1 \ \text{TeV}$, decays into twin vector bosons through mass mixing begin to dominate, and there is no favourable asymmetry in the branching fractions, as discussed in this section. Even at higher masses, the effects of heavy Higgs decays into light Higgs do not compensate sufficiently, as this partial width scales with $m_X$ in the same way as the partial width for longitudinally polarised weak bosons.

The constraint on the decay width from the required reheating temperature (\ref{ReheatRange}) translates into a constraint on the size of the coupling $\lambda_x x$. For $m_X\gtrsim m_h$, this gives $10^{-8.5}\, \text{GeV} \lesssim \lambda_x x\lesssim 10^{-6}$ GeV, while for lower masses, this range increases to $10^{-7}\, \text{GeV} \lesssim \lambda_x x\lesssim 10^{-5.5}$ GeV at $m_X\sim 20$ GeV.

The gray bands in Figure \ref{Fig:decayWidths} highlight regions where our analytic estimates of the partial widths encounter enhanced uncertainties arising from the bottom and charm thresholds in both sectors. Over most of these ranges, we estimate the size of these uncertainties to be either $\sim 10\%$ or confined to very small subregions. The thicknesses of these bands have been chosen conservatively, and ultimately the branching ratios should be accurate to within a factor of $\pm \Lambda_{QCD}$ of the bottom and charm mass thresholds. In particular, the prescription of \cite{Drees:1989du} has been followed for approximating the bottom partial width close to the open flavour threshold. Resonant decay into gluons from bottomonia mixing has been neglected, although these resonant mass ranges are expected to be only $\sim$ MeV wide at the CP-even, spin-0 bottomonia masses $m_X=m_{{\chi_b}_i}$ (see \cite{Drees:1989du} and \cite{Baumgart:2012pj}). It should be noted, however, that at temperatures above that of the QCD phase transition, the quark decay products behave differently compared to that expected in a low temperature environment. In particular, for hot enough temperatures, the $b$ or $c$ quarks may not hadronise and the partonic partial widths may more reliable. The applicability of the treatment of the flavour thresholds used here may therefore not be valid if the decay occurs in the hot early universe. However, it is only very close to the threshold itself (within several GeV) that this uncertainty becomes significant. Finally, quark masses have been neglected in the gluon partial width. For $m_X$ close to the flavour thresholds, this approximation breaks down, but the gluon branching fraction is only $\sim 10\%$ and so the error does not contribute to the uncertainty of the total width by more than this order (it is this uncertainty that is responsible for most of the extension of the length of the gray bands about the flavour threshold).

Close to the charm threshold, the analogous uncertainties are even more poorly understood. Below the charm threshold, hadronic decays of a light scalar are highly uncertain (see \cite{Clarke:2013aya} for discussion). We avoid these regions altogether by restricting our considerations to $m_X$ roughly above the twin charm threshold. Note that below the SM charm threshold, the smaller decay rate of a Higgs-like scalar necessitates larger couplings $\lambda_X x$ for $X$ to have a lifetime within the required reheating window. The larger couplings then imply potentially stronger constraints from invisible mesonic decays. See \cite{Baumgart:2012pj, Clarke:2013aya, Haisch:2016hzu} for further discussion and recent analysis of the pertinent experimental constraints.

Taken together, the results in Figures \ref{Fig:decayWidths} and \ref{Fig:brRatio} bear out the expectation that a scalar $X$ with symmetric couplings to the Standard Model and twin sectors may nonetheless inherit a large asymmetry in partial widths from the hierarchy between the scales $v$ and $f$. Across a wide range of masses $m_X$, the asymmetry is proportional to (or greater than) $v^2/f^2$, tying the reheating of the two sectors to the hierarchy of scales.

Before proceeding to our computation of cosmological observables, we comment on an alternative variation on the reheating mechanism presented here that involves having $X$ odd under the twin parity. This permits two renormalisable interactions with the Higgses to give a Higgs potential of the form:
\bea
\mathcal{V} \supset m_0^2 \left( |H_A|^2 + |H_B|^2 \right) + \lambda_0 \left( |H_A|^4 + |H_B|^4 \right) + \epsilon X^2 \left( |H_A|^2 + |H_B|^2 \right) + \tilde{\epsilon} X \left( |H_A|^2 - |H_B|^2 \right).\label{OddPot}
\eea
If $X$ then acquires a vev at some scale, it may be possible to arrange for the resulting spontaneous breaking of the $\mathbb{Z}_2$ to give that required in the Higgs potential. However, we find that, in order for $X$ to be long-lived and reheat the universe, its couplings to the Higgs must be highly suppressed and therefore that the resulting vev of $X$ required to explain the soft $\mathbb{Z}_2$-breaking in the Higgs potential must be many orders of magnitude above the twin scale. If this is to be identified with the characteristic mass scale of $X$, then a UV-completion of the twin Higgs is required for anything further to be said of the prospects of this possibility. However, if such a UV completion has similar structure to the couplings in (\ref{OddPot}), then asymmetric reheating may require a cancellation between the odd and even couplings of $X$ to the Higgs potential in order to suppress its twin-sector branching fraction (because the odd coupling appears with opposite signs in the coupling between $X$ and the $h_A$ and $h_B$ states). We do not consider this possibility further.

\subsection{Imprints on the CMB} \label{Sec:CMBfx}

For appropriate values of $m_X$, the out-of-equilibrium decay of $X$ reheats the two sectors to different temperatures and effectively dilutes the energy density in the twin sector. We obtain an analytic estimate of the effects of the $X$ decay on the number of light degrees of freedom observed from the CMB by approximating both the decay of $X$ and the decoupling of species as instantaneous in Section \ref{Sec:instant}. We then demonstrate that this estimate is reliable over most of the parameter space of interest with a numerical calculation in Section \ref{Sec:cosmosim}. In Section \ref{Sec:neutrinos} we consider neutrino masses and their joint constraints with $N_{\rm eff}$.


\subsubsection{Analytic estimate of $N_{\rm eff}$} \label{Sec:instant}

If $X$ dominates the energy density of the universe and then decays, depositing energy $\rho_\text{SM}$ and $\rho_\text{t}$ into the SM and twin sectors respectively, then the temperature ratio is determined by 
\begin{equation}
\frac{\rho_\text{t}}{{\rho_\text{SM}}} = \frac{g^\text{t}_{\star}(T^\text{t}_{\rm reheat})}{g^\text{SM}_{\star}(T^\text{SM}_{\rm reheat})} \left( \frac{T^\text{t}_{\rm reheat}}{T^\text{SM}_{\rm reheat}}\right)^4 \approx \frac{\Gamma(X \rightarrow \text{Twin})}{\Gamma(X \rightarrow \text{SM})},\label{EnergyRatio}
\end{equation}
where $T^\text{SM}_{\rm reheat}$ and $T^\text{t}_{\rm reheat}$ are the reheating temperatures for each sector, while $g^\text{SM}_\star$ and $g^\text{t}_\star$ are the SM and twin effective number of relativistic degrees of freedom, respectively. We have assumed that the two sectors are cool enough that they have already decoupled. We point out that not only does the number of effective degrees of freedom in each sector need to be evaluated at the temperature of that sector, but that $g^\text{t}_\star$ and $g^\text{SM}_\star$ differ as functions of temperature due to the differences in the spectra of the sectors, as seen in Figure \ref{Fig:geff}. As is well-known \cite{Kolb:1990vq}, reheating is a protracted process that occurs over a time-scale given by the lifetime of the reheaton. During this time, the temperature of the plasma cools slowly because, while the energy is being replenished by the decay of the reheaton, it is simultaneously diluted and redshifted with the expansion of the universe. It is assumed in (\ref{EnergyRatio}) that any primordial energy density in either sector is subdominant.

The temperatures of both sectors then redshift in the same way, so the only additional differences between their temperatures arise from changes to the effective number of degrees of freedom in each sector. By conservation of comoving entropy within each sector, each evolves as $T^i_{eq}/T^i_\text{reheat} = \left( g^i_{\star}(T^i_\text{reheat})/g^i_{\star}(T^i_{eq}) \right)^{1/3}a(T_\text{reheat})/a(T_{eq})$ where $T^i_{eq}$ is the temperature of the sector at matter-radiation equality, which the CMB probes as explained in Section \ref{Sec:Limits}, and $a(T)$ is the scale factor as a function of temperature. In the mirror Twin Higgs model, the two sectors have the same number of light degrees of freedom at recombination (three neutrinos and a photon, assuming that the neutrinos are still relativistic), so 
\begin{equation}
\left(\frac{T^\text{t}_{eq}}{T^\text{SM}_{eq}}\right)^4 = \left(\frac{T^\text{t}_{\rm reheat}}{T^\text{SM}_{\rm reheat}}\right)^4 \left(\frac{g^\text{t}_{\star}(T^\text{t}_{\rm reheat})}{g^\text{SM}_{\star}({T}_{\rm reheat})}\right)^{4/3}= \frac{\Gamma(X \rightarrow \text{Twin})}{\Gamma(X \rightarrow \text{SM})} \left(\frac{g^\text{t}_{\star}(T^\text{t}_{\rm reheat})}{g^\text{SM}_{\star}({T}_{\rm reheat})}\right)^{1/3}.\label{reheat}
\end{equation}
As our range of reheat temperatures encompasses the QCD phase transitions of both sectors, the factors of $g_{\star}$ can be important.

Given the temperatures of the two sectors after $X$ decays, we can obtain a simple estimate of the contribution to $N_{\rm eff}$ that neglects the impact of masses of the twin neutrinos discussed in Section (\ref{Sec:Neff}), 
\begin{align} 
(\Delta N_{\rm eff})_{m_\nu = 0} &= \frac{4}{7} \left(\frac{11}{4}\right)^{4/3} g^\text{SM}_{\star}(T^\text{SM}_{eq}) \frac{\rho_\text{t}(T^\text{t}_{eq})}{\rho_\text{SM}(T^\text{SM}_{eq})} \\
&\approx 7.4 \times \frac{\text{Br}(X \rightarrow \text{Twin})}{\text{Br}(X \rightarrow \text{SM})} \left(\frac{g^\text{t}_{\star}(T^\text{t}_{\rm reheat})}{g^\text{SM}_{\star}(T^\text{SM}_{\rm reheat})}\right)^{1/3}.
\end{align}
In this limit the most recent Planck data give a $2\sigma$ bound of $\Delta N_{\rm eff} \lesssim 0.40$ assuming pure $\Lambda$CDM+$N_{\rm eff}$ \cite{Ade:2015xua}. This translates into the requirement $\frac{\rho_\text{t}(T^\text{t}_{eq})}{\rho_\text{SM}(T^\text{SM}_{eq})} \approx \frac{\Gamma(X \rightarrow \text{Twin})}{\Gamma(X \rightarrow \text{SM})}  \lesssim 0.05$, ignoring possible differences in $g_{\star}$. 

Of course, as discussed in Section \ref{sec:thermal}, the twin neutrino masses are relevant at the temperature of matter-radiation equality, so we can obtain a more meaningful estimate of $\Delta N_{\rm eff}$ using the results of Section \ref{Sec:Neff} evaluated at the twin temperature determined above:
\begin{eqnarray} \label{Eq:deltaNeff}
\Delta N_{\rm eff} &=& \left(\frac{11}{4}\right)^{4/3} \frac{120}{7 \pi^2 \left(T^\text{SM}_{eq}\right)^4} \left( \rho^\text{t}_\gamma\left(T^\text{t}_{eq}\right) + \sum_\alpha 3 w^t_{\nu_\alpha}\left(T^\text{t}_{eq}\right) \rho^\text{t}_{\nu_\alpha}\left(T^\text{t}_{eq}\right)\right) \\
T^\text{t}_{eq} &=& T^\text{SM}_{eq} \left(\frac{\Gamma(X \rightarrow \text{Twin})}{\Gamma(X \rightarrow \text{SM})}\right)^{1/4} \left(\frac{g^\text{t}_{\star}(T^\text{t}_{\rm reheat})}{g^\text{SM}_{\star}({T}^\text{SM}_{\rm reheat})}\right)^{1/12}\label{TwinReh}
\end{eqnarray}
with $T^\text{SM}_{eq} \approx 0.77 \text{ eV}$ \cite{Ade:2015xua} the photon temperature. While the right-hand side of this equality has implicit dependence on  $T^\text{t}_{eq}$ through $g^\text{t}_{\star}$, this is only important if the reheating occurs between the SM and twin QCDPTs and the neglecting of the factors of $g_\star$ is otherwise reliable. With the further inclusion of Standard Model neutrino masses or an extra sterile neutrino, the bound described above weakens to $\Delta N_{\rm eff} \lesssim 0.7$. As discussed in Section \ref{Bounds}, we are not aware of any analyses specific to our model involving both pure dark radiation and three sterile neutrinos with masses of order the photon decoupling temperature of the CMB and possibly cooler temperatures. In the absence of such an analysis, we use the inequality $\Delta N_{\rm eff} \lesssim 0.7$ to indicate where the present CMB measurements are likely to constrain the light degrees of freedom of this model, leaving a more detailed analysis of the CMB constraints as future work. In this case, the bound on the decay width ratio is $\frac{\Gamma(X \rightarrow \text{Twin})}{\Gamma(X \rightarrow \text{SM})}  \lesssim 0.09$. The next generation of CMB experiments are projected to strengthen this constraint to $\Delta N_{\rm eff} \lesssim 0.02$ at the $1\sigma$ level \cite{Abazajian:2013oma}.

\subsubsection{Numerical Calculation of $N_{\rm eff}$} \label{Sec:cosmosim}

A more precise study of the effect of $X$ decay on the number of effective neutrino species at recombination may be performed by numerically solving a system of differential equations for the entropy in $X$ and the two sectors as a function of time. Following the analysis of Chapter 5.3 of \cite{Kolb:1990vq} we have
\begin{gather}
H=\frac{1}{a}\frac{da}{dt}=\sqrt{\frac{1}{3M_{Pl}^2}(\rho_X+\rho_{SM}+\rho_t)}\\
\frac{d\rho_X}{dt}+3H\rho_X=-\Gamma_X\rho_X \label{Eq:XEnDec}\\
\rho_i=\frac{3}{4}\left(\frac{45}{2\pi^2 g^i_{\star}}\right)^{1/3}S_i^{4/3}a^{-4}\\
S_i^{1/3}\frac{dS_i}{dt}=\left(\frac{2\pi^2 g^i_{\star}}{45}\right)^{1/3}a^4\Big(\rho_X\Gamma_{X\rightarrow i}+\frac{dq_{j\rightarrow i}}{dt}\Big),\label{Eq:XHeat}
\end{gather} 
where $S_i$ are comoving entropy densities and it has been assumed that $X$ is cold by the time it decays so that $\rho_X = m_X n_X$ with number density $n_X$ (this is reliable as we only consider $m_X>10$ GeV, which is above the decoupling temperature of $\sim 1$ GeV). The rate of heat flow from sector $j$ to $i$ per proper volume, $\frac{dq_{j\rightarrow j}}{dt}$, is defined in (\ref{HeatFlow}). To account for the temperature-dependence of the effective number of relativistic degrees of freedom in each sector, these equations are solved iteratively in the profiles of $g^i_\star(T^i)$. 

The equations are solved in three stages: before, during and after the decoupling of the SM and twin sectors. The ratio $f/v$ is fixed to $4$ for this analysis. Initial conditions were chosen with $\rho=10^{-12}\rho_X$, for combined SM and twin energy densities $\rho$. However, it is only the requirement that the initial energy density of $X$ dominates over that of the SM and twin sectors that is important for simulating the cosmology over the times of interest here, as the entirety of the latter is then generated by the subsequent decay. The results close to the decoupling and reheating epochs are otherwise insensitive to the initial conditions and ultimately match onto the standard outcome \cite{Kolb:1990vq} expected by equating the Hubble rate with the decay rate of X. The sectors are assumed to be in thermal equilibrium and sharing entropy until a temperature of $10$ GeV, below which they are evolved separately with the heat flows $\frac{dq_{i\rightarrow j}}{dt}$ switched on. Elastic scatterings were neglected from the heat flow rate to accelerate the computation. It was verified for the results found below that their contribution to the heat flow was always $\lesssim 10\%$ while the heat flow was itself not dominated by the Hubble rate. Heat flow was switched off again once the twin temperature reaches $0.1$ GeV, by which time thermal decoupling is long-since complete, and the sectors are subsequently evolved separately. Again, although the strengthening of the colour force and the QCDPT make the perturbative tree-level computation of the scattering rates unreliable at temperatures below $\sim 1$ GeV, as found in Section \ref{Sec:Decoupling} and also in the results below, the sectors decouple above these temperatures. Notably, the impact of $X$ on the expansion rate causes decoupling to occur at slightly hotter temperatures than expected from the analysis of Section \ref{Sec:Decoupling} for the decoupling in the standard cosmology.


\begin{figure}[h!]
\centering
\includegraphics[width=0.8\linewidth]{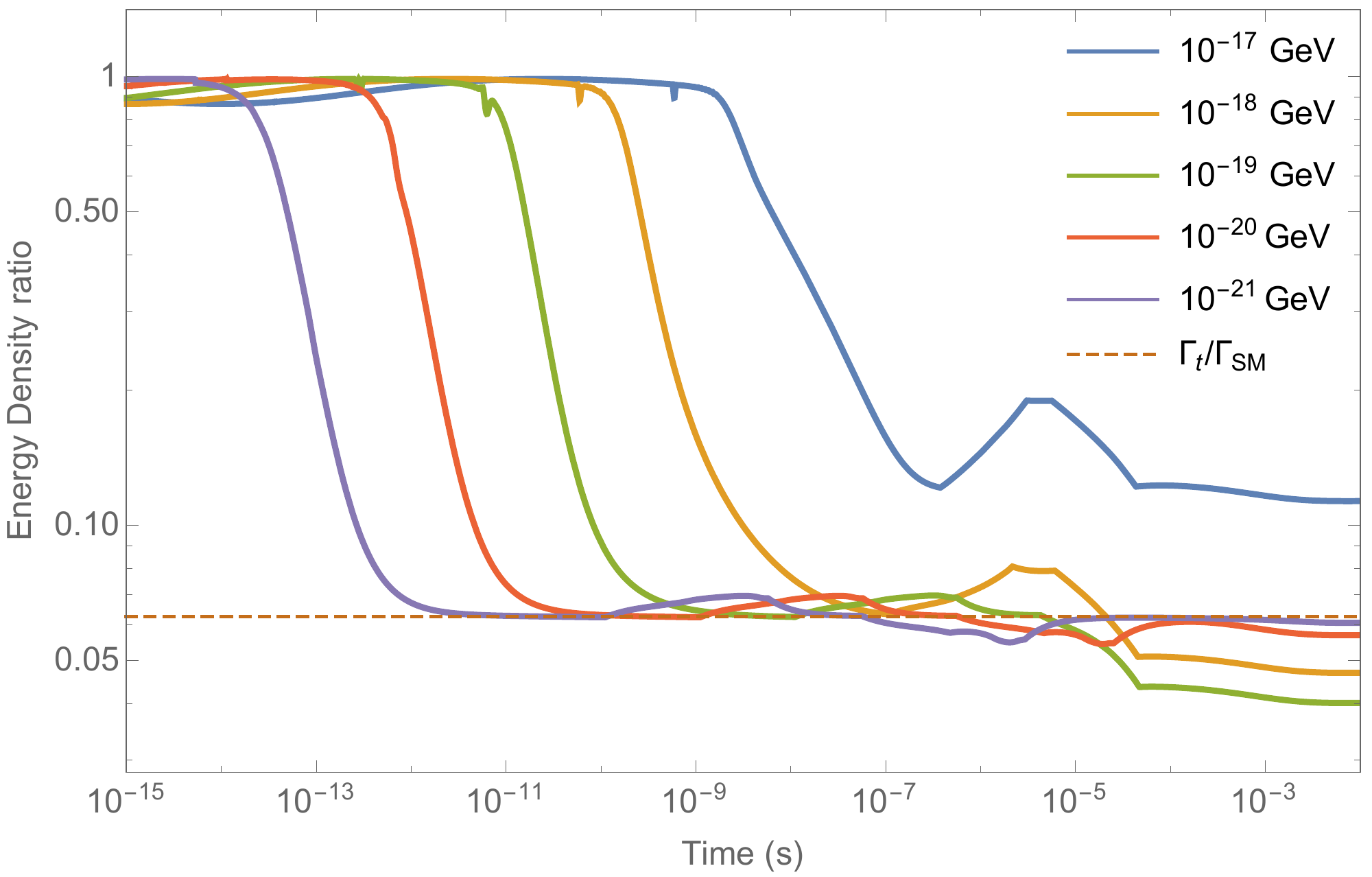}
\caption{Ratio of twin to SM energy densities throughout decoupling and reheating, for different decay rates $\Gamma_X$. The dashed line corresponds to the prediction of from the ratio of decay widths, here selected to be $1/16$.}
\label{Fig:enRatio}
\end{figure} 

The ratio of energy densities in each sector determines $N_\text{eff}$, from (\ref{Eq:deltaNeff}). A plot of this ratio over time is shown in Figure \ref{Fig:enRatio}, with the expectation under the approximations of the previous section shown as well. 
This approximation is reliable as long as the lifetime of $X$ is much longer than the temperature at which decoupling concludes, here $\sim 1$ GeV. The larger asymptotic value of the ratio of the blue line arises because the lifetime lies close to the decoupling period, so that a significant fraction of the energy is transferred while the sectors are thermalised or partially thermalised and does not contribute toward asymmetric reheating. Equivalently, as will be discussed below, insufficient time elapses between decoupling and reheating for the twin energy density to dilute and be repopulated by the decays to the level predicted by (\ref{EnergyRatio}). The subsequent bump represents the period between the reheating of the twin sector by its QCD phase transition followed by that of the SM. The green and orange lines correspond to reheating temperatures that lie between SM and twin QCD phase transitions. In these cases, the reheating of the SM from the subsequent SM QCD phase transition raises its energy density relative to the twin sector above that expected from the ratio of branching fractions. As this occurs after the lifetime of the reheaton, the estimate of the reheating temperatures presented in (\ref{reheat}) is still good as subsequent changes in the ratio due to the evolution of $g_{\star}$ are accounted for in our analysis of the reheating scenarios. 


%
The steep drop in the energy density ratio corresponds to the brief period during which the energy density of the twin sector present at decoupling dilutes and redshifts, which continues until it reaches a comparable size to the energy density that is being replenished by reheating. If the twin-sector branching fraction is highly suppressed, as can occur in the ``valley'' region in Figure \ref{Fig:brRatio} with $m_h\lesssim m_X\lesssim 2m_h$, then a longer time is required for this to happen, especially close to the decay epoch where the diminishing of the $X$ population also contributes to a reduced reheating rate. These effects can prolong the time required for the energy density ratio to converge to the asymptotic prediction of (\ref{EnergyRatio}).

\begin{figure}[h!]
\centering
\includegraphics[width=0.8\linewidth]{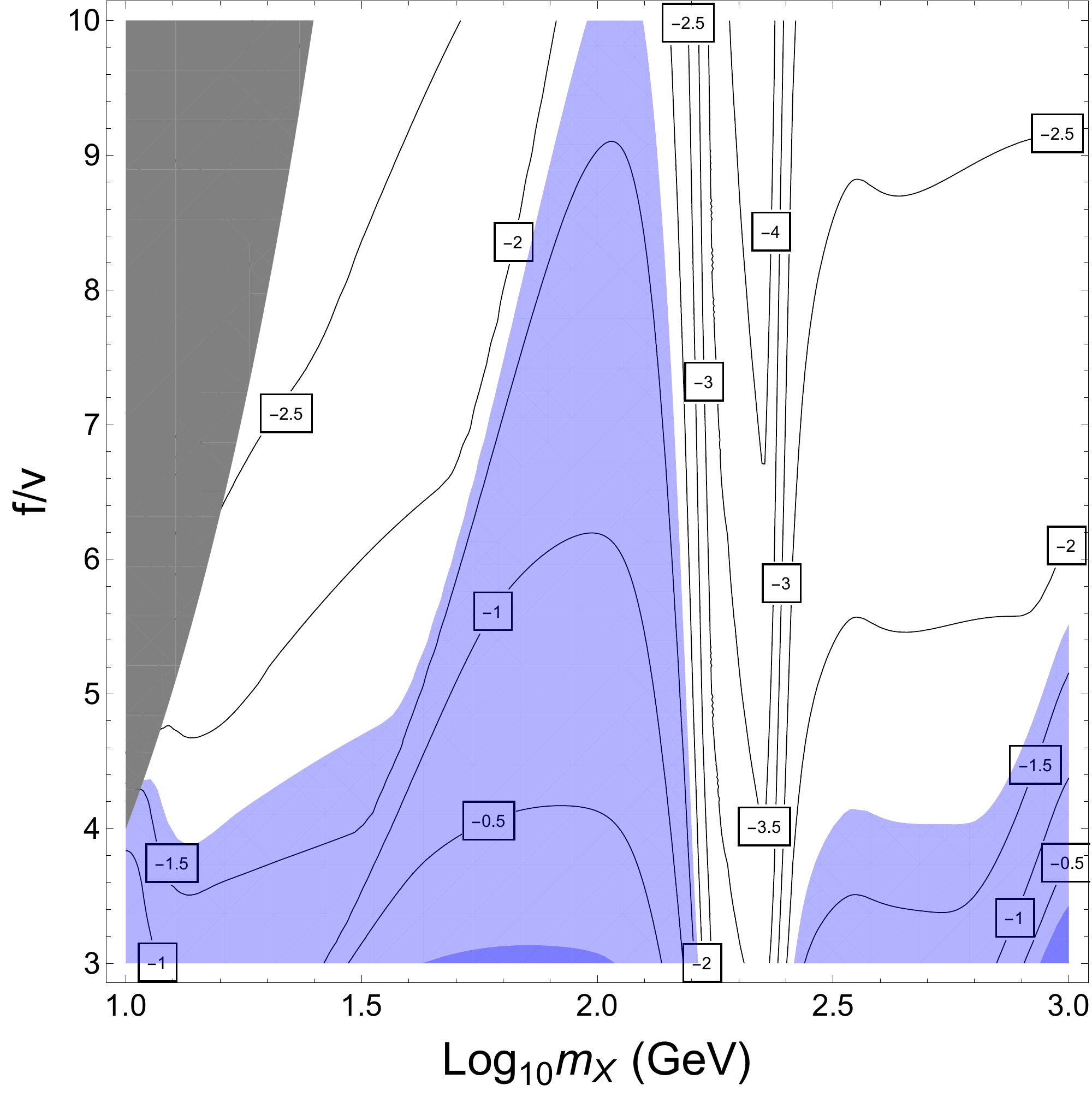}
\caption{Contours of $\log_{10}\Delta N_{\rm eff}$ as a function of $m_X$ and $f/v$, for $T^{\text{SM}}_{\rm reheat}=0.7$ GeV. 
The dark blue region is in tension with Planck, while the light blue region will be tested by CMB-S4. Gray regions are where the $X$ mass is below the twin charm threshold and our calculation of the twin sector partial width is unreliable.}
\label{Fig:XNeffmin}
\end{figure}

Contour plots of $\Delta N_{\rm eff}$ as a function of $m_X$ and $f/v$ appear in Figure \ref{Fig:XNeffmin}, along with current and predicted bounds using the analytic results of Section \ref{Sec:instant}. The minimum neutrino mass configuration with Dirac masses has also been assumed, although the results are relatively insensitive to this provided that the twin neutrino masses are not well above the eV scale. A SM reheating temperature of $0.7$ GeV has been assumed. At this temperature, we have verified using the numerical calculation of Section \ref{Sec:cosmosim} that the twin sector reheating temperature is always roughly above the twin neutrino decoupling temperature over the parameter space of the figure, ensuring that the neutrinos thermalise once produced in the decays and hence that the predictions of Section \ref{Sec:instant} are valid. A treatment of the case in which the twin neutrinos are produced below their decoupling temperature is beyond the scope of this analysis, but would involve the computation of the phase space spectrum of the neutrino decay products of the $X$. 

Also, as discussed in Section \ref{Sec:Decoupling}, a large temperature difference may partially relax back if reheating occurs close to sector decoupling. However, a reliable calculation of the heat flow at the temperatures of interest here must incorporate non-perturbative effects. We do not perform such a computation, but note that, at a slightly higher SM reheating temperature of $2$ GeV where this computation is more reliable, $\Delta N_{\rm eff}$ in Figure \ref{Fig:XNeffmin} can be raised by up to an order of magnitude in the region with $f/v\lesssim 4$ and $150\,\text{GeV}\lesssim m_X\lesssim 200\,\text{GeV}$, notably where the twin sector partial width is suppressed relative to the SM by several orders of magnitude. The resulting $\Delta N_{\rm eff}$ prediction is, nevertheless, still out of observable reach. At the lower SM reheating temperature assumed in Figure \ref{Fig:XNeffmin}, it is expected that decoupling will be further advanced and the enhancement in $\Delta N_{\rm eff}$ would be weaker.


We emphasize that, if the lifetime of $X$ is sufficiently close to the time of decoupling, or equivalently, that the reheating temperature is sufficiently close to the decoupling temperature, then the residual twin energy density left-over may be comparable to or greater than that regenerated by reheating. Consequently, the suppression in $\Delta N_{\rm eff}$ would be less than that predicted in (\ref{reheat}). In this respect, the projection of Figure \ref{Fig:XNeffmin} should be regarded as a lower bound on $\Delta N_{\rm eff}$. In the regions of high suppression, such as the ``valley'' region, the full asymmetry may not be generated before the complete decay of $X$ when the reheating temperature is of similar order as the decoupling temperature. In particular, for the reheating temperature chosen here of $0.7$ GeV and branching fraction ${\rm Br}(X\rightarrow \text{Twin})\sim 10^{-5}$, the numerical calculation of the energy density ratio saturates at $\sim 4\times 10^{-5}$. We do not include this effect in Figure \ref{Fig:XNeffmin} as its only impact is to mildly shift the unobservably small $\Delta N_{\rm eff}=10^{-4}$ contour. Lower reheating temperatures would agree with the prediction of (\ref{EnergyRatio}) were it not for the caveat that the twin neutrinos may be produced out of equilibrium. However, this minimum value at which $\Delta N_{\rm eff}$ is saturated can grow significantly with hotter reheating temperatures upon which it is highly dependent. 

CMB-S4 observations will be able to probe a large portion of the most natural parameter space, save the region $m_h \lesssim m_X \lesssim 2 m_h$ where decays into the Standard Model dominate well beyond the ratio $f^2/v^2$, as previously discussed. Significantly, precision Higgs coupling measurements at the LHC are unlikely to probe the mirror Twin Higgs model beyond $f \sim 4 v$, so that the observation of additional dark radiation may be the {\it first} signature of a mirror Twin Higgs.

\subsubsection{Neutrino Masses} \label{Sec:neutrinos}

In addition to the bounds on $N_{\rm eff}$, we must also respect the bounds on neutrino masses. The analysis remains nearly the same as in Section \ref{Sec:meff}, but now with the twin neutrinos at a lower temperature, as determined above. As mentioned above, for large enough $f/v$ and SM reheating temperature sufficiently close to the lower bound, the reheating temperature of the twin sector may be below the twin neutrino decoupling temperature and the resulting energy density would be more difficult to compute. For simplicity, we choose $\lambda_x x$ large enough such that the twin reheating temperature is always above the twin neutrino decoupling temperature. 

As before, we compute $m_\nu^{\rm eff}$ as 
\begin{equation}
m_\nu^{\rm eff} = \frac{n^\text{t}_\nu}{n^\text{SM}_\nu} \sum_\alpha m^\text{t}_{\nu_\alpha}.
\end{equation}  
In relating the scale factors at neutrino decoupling in each sector, we now have to use the above temperature ratio to find, analogously to Section \ref{Sec:meff}, that 
%
%
\begin{equation}
m_\nu^{\rm eff} = \left( \frac{\Gamma_\text{t}}{\Gamma_\text{SM}} \right)^{3/4}  \left( \frac{g^\text{t}_\star\left(T^\text{t}_\text{reheat}\right)}{g^\text{SM}_\star\left(T^\text{SM}_\text{reheat}\right)}\right)^{1/4} \left(\frac{f}{v}\right)^n \sum_\alpha m^\text{SM}_{\nu_\alpha},
\end{equation}
where, again, $n = 1$ for Dirac masses and $n = 2$ for Majorana masses. Interestingly, if the branching ratios scale as $\Gamma_\text{t}/\Gamma_\text{SM} = (v/f)^2$, then we have $m_\nu^{\rm eff} \propto (f/v)^{-3/2 + n}$, so the contribution grows with $f/v$ for Majorana masses, but is suppressed for Dirac masses.

\begin{figure}[h!]
\centering
\includegraphics[width=0.9\linewidth]{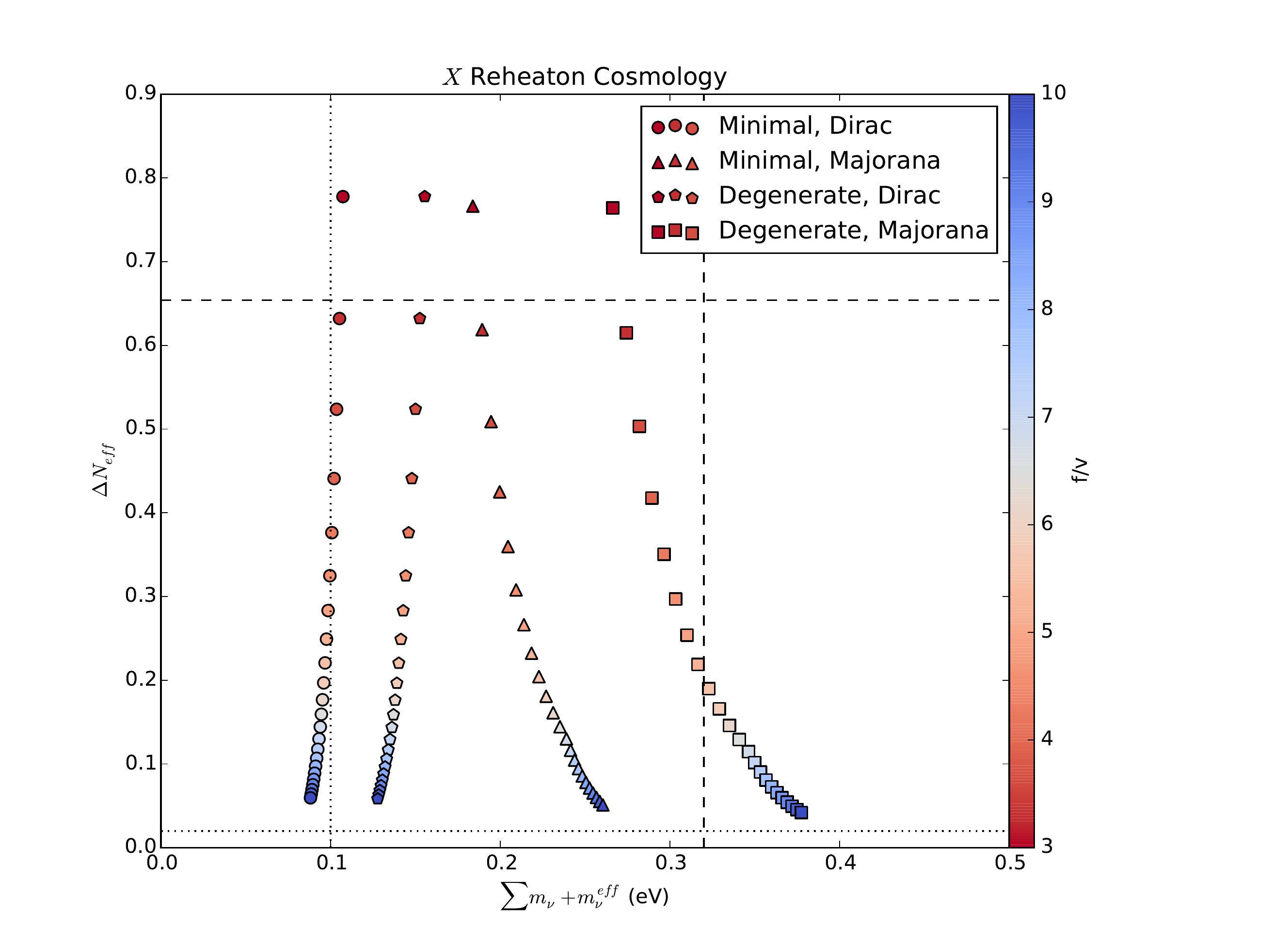}
\caption{Predicted values of $\Delta N_{\rm eff}$ and $\sum m_\nu + m_\nu^{\rm eff}$ for minimal and degenerate neutrino mass spectra with both Dirac and Majorana masses for $f/v$ from 3 to 10. The Planck 2015 \cite{Ade:2015xua} bounds on $\sum m_\nu$ and $N_{\rm eff}$, as discussed in Section \ref{Bounds}, are represented by the dashed lines, and the projected CMB-S4 constraints are given by the dotted lines. It has been assumed that $\frac{\Gamma_\text{t}}{\Gamma_\text{SM}} \sim (v/f)^2$.  Note however, that, from Figure \ref{Fig:XNeffmin}, this scaling of the partial widths holds only for the mass range $50\,\text{GeV}\,\lesssim m_X\lesssim 120\,\text{GeV}$, outside of which the twin partial width is more suppressed and the model is only testable through $\Delta N_{\rm eff}$ over a smaller range in $f/v$.}
\label{Fig:asymPred}
\end{figure}

As before, we consider the minimal mass spectrum of $m_\nu = \left[0.0,0.009,0.06 \text{ eV}\right]$ and a degenerate spectrum of $m_\nu = \left[0.1\text{ eV} ,0.1\text{ eV},0.1 \text{ eV}\right]/3$. In Figure \ref{Fig:asymPred} we plot the predictions of the $X$ reheating for $\Delta N_{\rm eff}$ and $m_\nu^{\rm eff}$ for both spectra and both Dirac and Majorana masses using the approximations of Section \ref{Sec:Limits}, for $f/v$ from 3 to 10 and assuming the $\frac{\Gamma_\text{t}}{\Gamma_\text{SM}} \sim (v/f)^2$ scaling; there are regions in the space of $m_X$ where the suppression of $m_\nu^{\rm eff}$ would be much higher. 

Dashed lines indicate the rough locations of present experimental limits from Planck 2015, and projected bounds from CMB-S4. As mentioned in Section \ref{Sec:meff}, we are unaware of any study of bounds on both $m_\nu^{\rm eff}$ and $\Delta N_{\rm eff}$ treated jointly. In the absence of this, we show present and projected constraints on $N_{\rm eff}$ and $\sum m_\nu$ from \cite {Ade:2015lrj} and \cite{CMB-S4:2016}, ignoring correlations, as described in Section \ref{Bounds}. 

\subsection{Thermal Production} \label{Eq:thermal}
In our discussion up to this point, we have been agnostic about the origin of the cosmic abundance of $X$ and have operated under the assumption that it absolutely dominates the cosmology before it decays. Here, we consider the possibility that $X$ was thermally produced through freeze-out and subsequently dominates the universe as a relic before decaying. This thermal history is viable, but places strong constraints on the mass and couplings of the $X$.

The energy density of relativistic species redshifts as $\rho_r \propto a^{-4} \propto T^4$, while the energy density of non-relativistic, chemically decoupled matter scales as $\rho_m \propto a^{-3}$. The energy density contained in the $X$ can therefore only grow relative to the energy density in the thermal bath once it becomes non-relativistic. We found in Section \ref{Sec:instant} that by recombination, $\rho_\text{t}/\rho_\text{SM} \lesssim 0.09$ is needed to evade current bounds on $\Delta N_\text{eff}$. Thus we need to have the energy density in the $X$ dominate over the SM and twin plasmas by more than this factor when it decays. 
If $X$ becomes non-relativistic instantaneously at the moment that its temperature reaches some fraction $c\sim \mathcal{O}(0.1)$ of its mass, then, as 
%
%
$T \propto 1/a$ and $\rho_X$ is $\sim 1/g_\star$ of the total energy density, the mass is required to satisfy $m_X\gtrsim 10/c\times g_\star\left(T = m_X\right)T^\text{SM}_{X \text{reheat}}$. 
Since the SM reheating temperature is strongly constrained to be above BBN, this effectively puts a lower limit on the mass of the $X$. Importantly, $X$ must freeze-out when relativistic or its energy density will be further Boltzmann suppressed. The lower limit on the mass of the $X$ becomes an upper limit on the $X$'s couplings - if it couples too strongly to the thermal bath, then it won't freeze out early enough to be hot.

%
%
In fact the situation is somewhat less favorable than the above analysis suggests, because it is relevant operators that must keep $X$ in thermal equilibrium. For an $X$ with the interactions introduced in Section \ref{Sec:reheaton}, the annihilations 
have rates that scale with temperature as $\Gamma \sim n_X \left\langle \sigma v \right \rangle \sim T$ for $T\gtrsim m_X,m_h$ (where $n_X$ is the number density of $X$ and $\left\langle \sigma v \right \rangle$ is its thermally averaged annihilation cross section). However, in a radiation-dominated universe, $H \sim T^2$. Thus, at high enough temperatures, $X$ is not in thermal equilibrium with the plasma and it is only once the universe cools enough that it may thermalise. 
Then, as the temperature drops, $XX \rightarrow q\bar{q}$ annihilations become suppressed by the Higgs mass and subsequent Boltzmann suppression causes $X$ to freeze-out. Note that the rates of these annihilation processes are controlled by the coupling $\lambda_x$, independently of $x$, which is unconstrained by itself (other processes mediated by $\lambda_x x$ are found to be subdominant in the ensuing analysis, for the range of $\lambda_x$ over which thermal production is successful). If the coupling is too weak to begin with, then the $X$ never thermalises and thermal production cannot happen. Thermal production therefore requires a careful balancing of parameters - small coupling $\lambda_x$ is preferred for $X$ to freeze-out hot and as early as possible, but the coupling is bounded from below by the requirement that $X$ reach thermal equilibrium. This combination of constraints severely restricts the size of the parameter space over which thermal production is viable to cases in which the coupling is selected so that $X$ enters and departs from thermal equilibrium at close to the same temperature. 

To obtain numerical predictions for this scenario, the calculation of Section \ref{Sec:cosmosim} was modified to account for the time after the freeze-out of $X$ before it becomes non-relativistic. During this period we use (\ref{Eq:nudis}) and (\ref{Eq:nuen}) for the energy density of the $X$, approximating decays as being negligible, before switching over to (\ref{Eq:XEnDec}) when the temperature drops below the mass of the $X$. The approximation that the $X$ does not decay appreciably while it is relativistic must be good if there is to be sufficient time for it to grow to dominate between becoming non-relativistic and decaying. The decay width of $X$ was fixed to $5\times10^{-21}\,\text{GeV}$, corresponding to a reheating temperature close to the $\sim 10$ MeV lower limit, in order to maximise the amount of time over which the energy density of $X$ may grow relative to the SM plasma, thereby providing the greatest possible reheating.

\begin{figure}[!h]
\centering
\includegraphics[width=.8\linewidth]{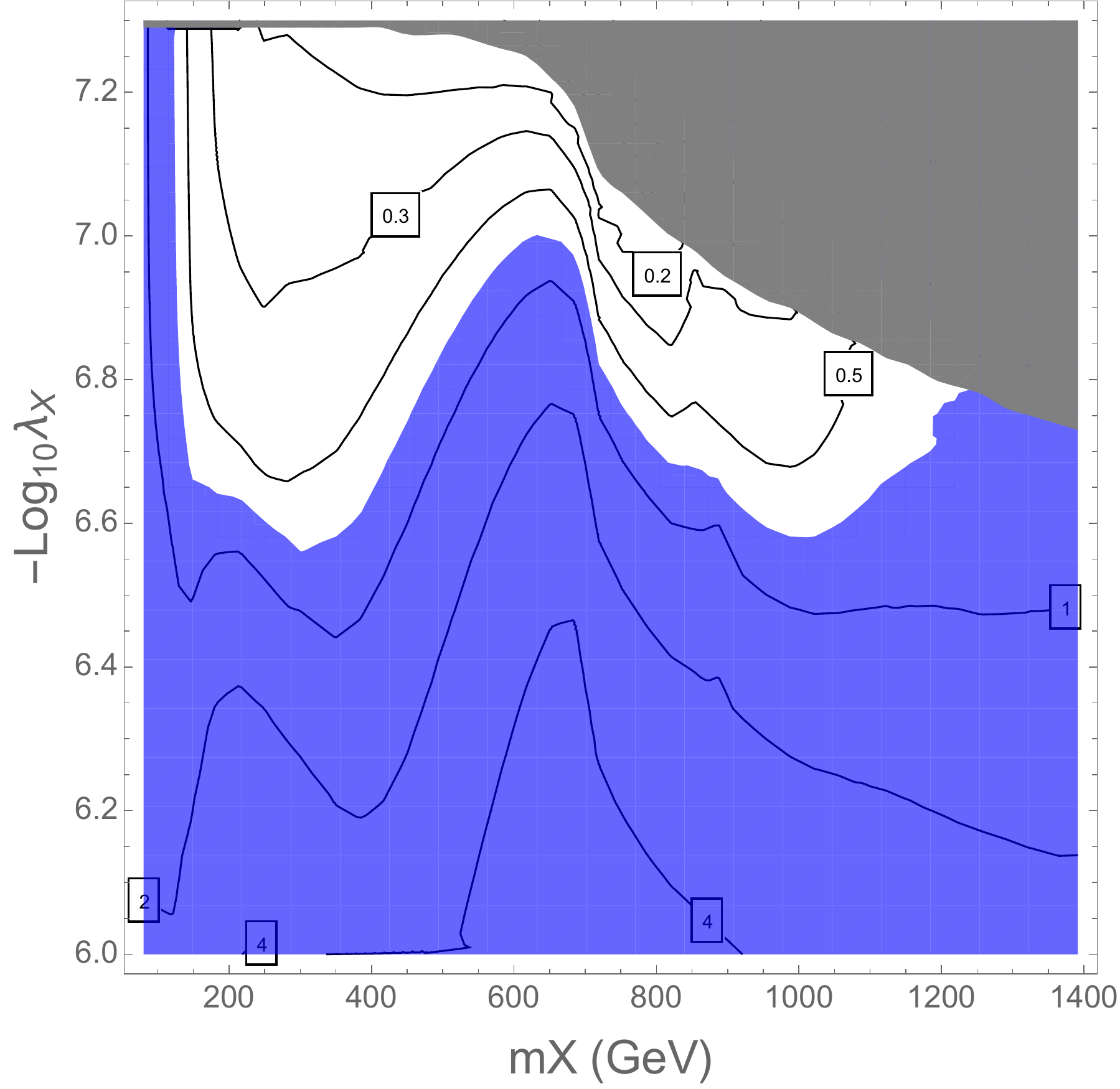}
\caption{Parameter space where thermal production of $X$ gives a large enough relic abundance to dilute the twin sector, for $f/v=4$. In the gray region, the coupling is too weak for $X$ to ever reach thermal equilibrium. The blue region is in tension with recent Planck measurements of $\Delta N_{\rm eff}$, whereas all of the white region will be tested by CMB-S4. Predictions presented here for $\Delta N_{\rm eff}$ close to the gray boundary are more uncertain because of the high sensitivity of the freeze-out temperatures to the coupling.}
\label{Fig:freezeout}
\end{figure}

The predictions for $\Delta N_\text{eff}$ from a thermally produced $X$ are shown in Figure \ref{Fig:freezeout} for the small regions of parameter space where this is viable, with $f/v=4$. We find that the dominant annihilation channels over this region are $XX\rightarrow t\bar{t}$ and $XX\rightarrow b\bar{b}$, mediated by the light Higgs, as well as their twin analogues, mediated by the heavy Higgs. As expected, the primordial energy density in the twin sector is too large compared to that generated by the $X$ for the asymmetric reheating to be effective when $m_X$ is too light ($\lesssim 100$ GeV in this case). Similarly, when the coupling is too strong, the X is held in equilibrium for longer and freezes-out underabundant compared to the twin energy density. However, when the coupling is too weak (the gray region), $X$ never thermalises to begin with (close to the boundary with this region, $X$ freezes-out almost immediately after thermalising). The peak in the contours occurs because of the ``$H$-funnel'' in which the twin Higgs resonantly enhances annihilations into twin quarks. All of this region will be testable by CMB-S4. 

\section{Twinflation} \label{sec:twinflation}


As an alternative to the model presented above of late, out-of-equilibrium decays of a $\mathbb{Z}_2$-symmetric scalar, one may imagine that the field driving primordial inflation reheats only the Standard Model to below the decoupling temperature of the two sectors. Production of the twin particles then ceases at some time after the temperature drops below the decoupling temperature during reheating.


To make this consistent with a softly-broken $\mathbb{Z}_2$ symmetry, we extend the inflationary sector and introduce a `twinflaton' that couples solely to the twin sector. The combined inflationary and twinflationary sectors respect the $\mathbb{Z}_2$ symmetry. However, if the two sectors are entirely symmetric then one generally expects both inflationary dynamics to happen coincidentally, which would result in identical reheating.
We therefore rely on soft $\mathbb{Z}_2$-breaking to give an asymmetry between the two sectors that causes the twinflationary sector to dominate the universe first. With the right arrangement we can end up with two distinct periods of inflation - a first caused by the ``twinflaton'' and a second that then reheats the Standard Model to below the decoupling temperature, having diluted the sources of twin-sector reheating from the first period.

One simple mechanism for $\mathbb{Z}_2$-breaking which is well-suited for introducing asymmetry to inflationary sectors is to introduce an additional $\mathbb{Z}_2$-odd scalar field $\eta$ (as was done in \cite{Berezhiani:1995am}). This admits linear and quadratic interactions to antisymmetric and symmetric combinations of the inflationary sector fields, respectively. When $\eta$ acquires a vev, this introduces an asymmetry in the fields to which it was coupled, dependent on the combination of its vev and its couplings. If $\eta$ is coupled to both the inflationary sectors and the Higgs sectors, it could be the sole source of $\mathbb{Z}_2$-breaking in a twinflationary theory. One may generally imagine that, in some UV completion, the mechanism that softly breaks the symmetry in the Higgs potential could also be the origin of the soft breaking of the inflationary sector.

Cosmologically, this possibility may have similar observational signatures as the model discussed in Section \ref{sec:late}, where the amount of twin-sector dark radiation is determined by the partial widths of the inflaton of the second inflationary epoch. If this dominantly couples to the SM, then $\Delta N_{\rm eff}$ will be suppressed which, while successfully resolving the cosmological problems of the Mirror Twin Higgs, may also be observationally inaccessible. However, additional, distinctly inflationary signatures may make this potentially testable by other cosmological observations.

The mechanism of twinflation completes a catalog of models of asymmetric reheating by late decays, which may be indexed by representations of the twin parity: the case of a $\mathbb{Z}_2$-even particle, in which a kinematic asymmetry in the partial widths provides the reheating asymmetry, the case of a $\mathbb{Z}_2$-odd particle, which can also provide the spontaneous $\mathbb{Z}_2$-breaking required in the Higgs potential, and the case where two distinct, long-lived particles couple to each sector, which may also be related to inflation.

\subsection{Toy Model}

As a toy model we here consider `twinning' the simple $\varphi^2$ chaotic inflation scenario. The inflationary dynamics in this case are easy to understand and we have the additional benefit that this inflationary model has been considered in the literature before as `Double Inflation' (see \cite{Silk:1986vc}, \cite{Polarski:1992dq} and \cite{Wands:2002bn}). We furthermore specialize to `double inflation with a break', where there are two distinct periods of inflation which produces a step in the power spectrum, and we consider the constraints that this places on our model. In this case, it is assumed that each inflaton field couples and therefore decays dominantly into the sector to which it belongs. 
We will comment briefly on the case without a break and the additional signals one could look for in that case. 

The potential of the inflationary sector for inflaton $\varphi_A$ and twinflaton $\varphi_B$ is
\begin{equation}
V = \frac{1}{2} m_A^2 \varphi_A^2 + \frac{1}{2} m_B^2 \varphi_B^2,
\end{equation} 
where $m_A \neq m_B$ may arise from soft $\mathbb{Z}_2$- breaking, perhaps related to the soft $\mathbb{Z}_2$-breaking in the Higgs potential. In order for the `twinflation' to occur first, we require that the energy of the $B$ field initially dominates the energy density of the universe. We take the initial positions of the fields to be the same and $m_B^2 \gg m_A^2$.\footnote{Note that merely giving the twin field a much larger initial condition does not instigate twinflation. The dynamics of the subdominant field in this case are such that it will track the dominant field and both will reach the critical value at the same time. This is easily confirmed numerically.} Call $\varphi_A(0) = \varphi_B(0) = n \sqrt{2} M_\text{pl} = n \varphi_c$, where $\varphi_c$ is the critical value at which inflation stops and $m_B = r m_A = r m$ with $n,r > 1$. The inflationary dynamics are then those of slowly-rolling scalar fields. At some point in the early universe we imagine that the slow-roll approximation holds for both fields and the inflationary sector dominates the universe. 
%
%
The dominating field then slow-rolls down its potential for $\frac{n^2-1}{2}$ $e$-folds, while the lighter field's velocity is suppressed by approximately $\frac{\varphi_A}{r^2 \varphi_B}$. Solving the system numerically reveals that the motion of $\varphi_A$ during this period can be neglected entirely.

After $\varphi_B$ reaches the critical value $\sqrt{2} \Mp$, it stops slow-rolling and begins oscillating around the minimum of its potential. For there to be two distinct periods of inflation, there must be a period where these oscillations dominate the universe, which requires that the energy densities of each inflaton $\rho_A$ and $\rho_B$ satisfy  $\rho_B(\varphi_c) = r^2 m^2 M_\text{pl}^2 \ > \ \rho_A(\varphi(0)) = n^2 m^2 M_\text{pl}^2 $ and therefore $r>n$. For a $\varphi_B^2$ potential, the energy in these oscillations redshifts as $\rho_B \sim a^{-3}$. Eventually, the energy density in $\varphi_B$ drops below that of $\varphi_A$ and a new epoch of inflation, driven by $\varphi_A$, begins. This provides a further $\frac{n^2 - 1}{2}$ $e$-folds of inflation to give $n^2 - 1$ in total, while the $B$-sector energy density is diluted away.

%
%


Note that in order for our toy model to reheat below the decoupling temperature of the two sectors, reheating must occur well after the end of inflation. If, during the coherent oscillation of an inflaton, it becomes the case that the inflaton decay width $\Gamma \sim H$, then reheating will occur and result in temperature $T_{\rm reheat} \sim 0.1 \sqrt{\Gamma M_\text{pl}}$. However, if $\Gamma \gg H$ when inflation ends, then all of the energy in the inflaton is immediately transferred and we instead have reheating temperature $T_{\rm reheat} \sim 0.1 \sqrt{m_\alpha M_\text{pl}}$ for an inflaton of mass $m_\alpha$. But in order for $T_{\rm reheat}\lesssim 1$ GeV, it is required that $m_\alpha \lesssim 10^{-7}$ eV, so this possibility that the inflaton is short lived is not viable.
The procedure of twinning inflationary potentials may be generalised to other, more realistic models, provided that this constraint upon the reheating temperature can be satisfied.


\subsection{Observability}

One could always make a twinflationary scenario consistent with observational constraints by letting the second inflationary period of inflation last long enough. In our toy model, this would correspond to setting $n$ high enough that the momentum modes which left the horizon during the first inflation have not yet re-entered the horizon - such a scenario would look exactly like single-field chaotic inflation.

Alternatively, we may also allow for $n$ small enough that all the momentum modes that left the horizon during the second inflation are currently sub-horizon. In this case, fluctuations at large enough wavenumbers (equivalently, small enough length scales) are `processed' (cross the horizon) at a different inflationary energy scale than those that were processed earlier, giving a step in the power spectrum. While Planck has measured the primordial power spectrum for modes with $10^{-4} \ \text{Mpc}^{-1} \lesssim k \lesssim 0.3 \ \text{Mpc}^{-1}$ (where the lower bound is set by the fact that smaller modes have not yet re-entered the horizon), proposed CMB-S4 experiments will increase this range \cite{CMB-S4:2016} somewhat, as will be discussed further below. We wish to show that the power spectrum of our toy model is not ruled out and, furthermore, may be observed in the coming decades. 

The height of the step in the primordial power spectrum is determined by the energy scale of each period of inflation, so modes crossing the horizon in the second inflationary period should be suppressed by a factor of $r^2 > n^2 \gtrsim 25$ compared to those exiting in the first period. This degree of suppression is ruled out by Planck for the range of modes over which it has reconstructed the power spectrum \cite{Ade:2015lrj}. A computation of the primordial power spectrum for double inflation was given in \cite{Polarski:1992dq}. It was found that significant damping does not occur for modes which cross outside the horizon during the first inflationary period, re-enter during the inter-inflationary period and again cross the horizon during the second inflationary period. It is only those scales which first cross the horizon during the second inflationary period that are significantly damped (although other features in the shape, such as oscillations, may be present for modes that are subhorizon during the intermediate period).



The relation of this characteristic scale to present-day observables is easily done using the framework given in \cite{Hong:2015oqa}. Let the subscripts $a,b,c,d,e$ respectively correspond to the beginning of the first inflationary period, the end of that period, the beginning of the second inflationary period, the end of that period, and the beginning of radiation domination. During the coherent oscillation periods, the inflaton acts as matter and the energy density falls as $\rho \propto a^{-3}$. Let $k_i$ be the momentum whose mode is horizon-size at the $i$ epoch; $k_i = a_i H_i$. 
\begin{figure}[h]
\centering
\includegraphics[width=1.0\linewidth]{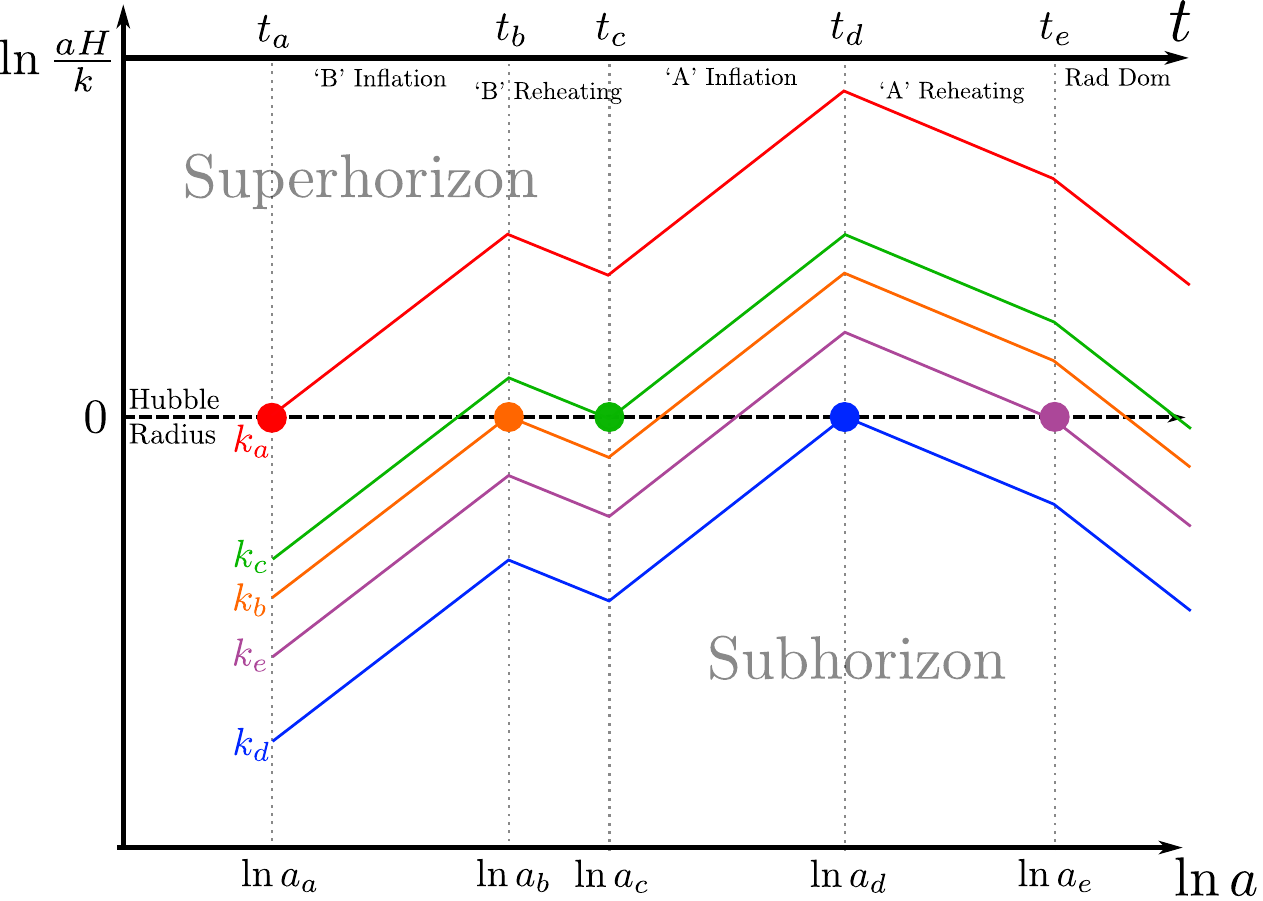}
\caption{Schematic evolution of the characteristic scales in Twinflation, as seen by comparing wavenumbers to the Hubble radius over time. 
Note that the time axis is not a linear scale.}
\label{Fig:scales}
\end{figure}
The scales $k_i$ can be related using the number of $e$-folds in each period, which are themselves determined from the first Friedmann equation. Denoting $N_{ij} = \ln \frac{a_j}{a_i}$, we have $k_a = e^{-N_{ab}} n k_b$, $k_b =e^{\frac{1}{2} N_{bc}} k_c$
%
%
and similarly for the other characteristic modes, where, in particular, slow-roll inflation predicts that $N_{ab} = N_{cd} = \frac{n^2-1}{2}$. 
%
%
The evolution of the characteristic momentum scales is shown schematically in Figure \ref{Fig:scales}. Finally, $k_e$ can be determined using the conservation of comoving entropy:
\begin{eqnarray}
k_e &=& \frac{\pi g_{\star}^{1/3}(T_0) g_\star^{1/6}(T_{\rm reheat}) T_0 T_{\rm reheat}}{3 \sqrt{10} M_\text{pl}},
\end{eqnarray}
where $T_0$ and $a_0$ are the temperature and scale factor today and $T_{\rm reheat}$ is the reheating temperature (which is sufficiently low that only SM particles are produced). We work explicitly with the convention $a_0 = 1$. The characteristic modes associated with the break can then be determined.

As mentioned above, \cite{Polarski:1992dq} shows that damping occurs for modes that exit the horizon only during the second inflationary period, so we should take the characteristic damping scale to be the smallest such scale, which here corresponds roughly to $k_b$
This can be determined as
\begin{eqnarray} \label{char}
k_b &=& n  e^{\frac{1}{2} N_{bc} - N_{cd} + \frac{1}{2} N_{de}} k_e \nonumber \\
&=& n \left(\frac{r}{n}\right)^{1/3} \exp\left(-\frac{n^2-1}{2}\right)  \left[ \frac{\frac{1}{2}m^2 M_\text{pl}^2}{\frac{\pi^2}{30}g_\star(T_{\rm reheat}) T_{\rm reheat}^4}\right]^{1/6} \frac{\pi g_{\star}^{1/3}(T_0) g_\star^{1/6}(T_{\rm reheat}) T_0 T_{\rm reheat}}{3 \sqrt{10}  M_\text{pl}}
\end{eqnarray} 
where $k_c$ only differs by the factor of $\left(r/n\right)^{1/3}$ (which is roughly close to unity). Once again, between $k_b$ and $k_c$ are oscillatory features, so $k_b$ should merely be taken as the rough characteristic scale of the damping.


\begin{figure}[h]
\centering
\includegraphics[width=1.0\linewidth]{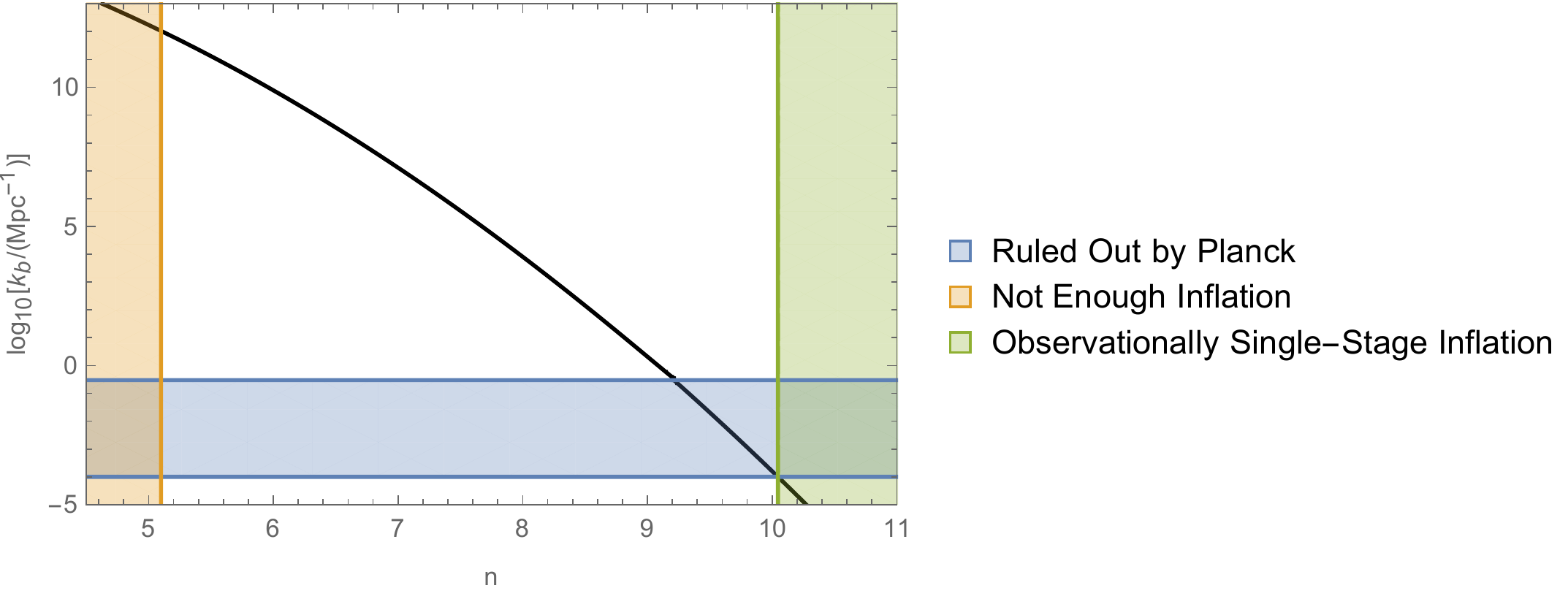}
\caption{The prediction for the characteristic suppression scale as a function of the initial values of the fields. The mapped regions should be interpreted not as having hard boundaries, but rather fuzzy endpoints where they break down. Here we have used $T_{\rm reheat} = 10 \ \text{MeV}$ and $r = 2n$.}
\label{Observable}
\end{figure}

Now the characteristic damping scale is determined by $m$, $n$, $r$, and $T_{\rm reheat}$. Our observational bound on $k_b$ is that Planck has not seen this suppression on momentum scales at which it has been able to reconstruct the primordial power spectrum from the angular temperature anisotropy power spectrum, which is roughly $k \lesssim 0.3 \ \text{Mpc}^{-1}$. We have constraints on the reheating temperature from rethermalization of the twin sector or interrupted big bang nucleosynthesis $10\, \text{MeV} \lesssim T_{\rm reheat} \lesssim 1 \ \text{GeV}$, on having a period of intermediate matter domination between the two inflations $r > n$ and on the total number of $e$-folds $n^2 - 1 \gtrsim 25$ to solve cosmological problems. Note that we require fewer $e$-folds of inflation than is typically assumed in the standard cosmology. Since the low reheating temperature gives fewer $e$-folds from reheating up to today, less inflation is needed to explain the large causal horizon and flatness.

The normalization of the spectrum provides a further constraint, the most recent measurement of which come from Planck \cite{Ade:2015lrj}. The scalar power spectrum at $k_\star = 0.05 \  \text{Mpc}^{-1}$ is measured to be $\mathcal{P}_{\mathcal{R}}(k_\star) = e^{3.094 \pm 0.034} \times 10^{-10}$. Then for $k_\star < k_c$ (i.e. $k_\star$ having left the horizon during the first period of inflation and not re-entered before the second, so no deviation from single-field inflation would be seen at this scale), the spectrum of \cite{Polarski:1992dq} yields the constraint 
\begin{equation} \label{norm}
2.03 \times 10^{-6} = \frac{r^2 m^2}{M_\text{pl}^2} \ln\left(\frac{k_b}{k_\star}\right)\left(\ln\frac{k_b}{k_\star} + \frac{n^2}{2}\right).
\end{equation}

The characteristic scale (\ref{char}) depends much more strongly on $n$ than it does on any of the other parameters. In Figure \ref{Observable}, we give a rough idea of the scale as a function of $n$, having set $T_{\rm reheat} = 10 \ \text{MeV}$ and $r = 2n$, while $m$ is chosen to satisfy the normalization condition. We also show the constraint on $k_b$ set by Planck. Note again that the region described as ``observationally single-stage inflation" \textit{does} still provide a solution to the problem of reconciling cosmology with the mirror Twin Higgs.

CMB-S4 will improve the constraint on $k_b$ through its improved measurement of polarization anisotropies \cite{CMB-S4:2016}. With only precision measurements of temperature anisotropies, the un-lensed power spectrum cannot be so easily reconstructed from the lensed spectrum. The effects of gravitational lensing of CMB place an upper limit on the size of primordial temperature anisotropies that can be measured \cite{Hu:2003vp}, which Planck has saturated. However, the polarization anisotropy power spectrum allows the removal of lensing noise from the temperature spectrum so that higher primordial modes can be detected. The polarization power spectrum itself also gives us another window into the high-$\ell$ modes of the primordial power spectrum, as the signal does not become dominated by polarized foreground sources until higher scales near $\ell \sim 5000$. CMB-S4 is projected to make cosmic variance limited measurements of both the temperature and polarization anisotropy power spectra up to the modes where they become foreground-contaminated and so provide additional information on the shape of the primordial power spectrum \cite{CMB-S4:2016}. The map from measurements of angular modes $\ell$ to contraints on spatial modes $k$ depends on the evolution of the power spectrum between inflation and the CMB, so forecasting constraints requires careful study. However, these improvements will not test most of the parameter space presented in Figure \ref{Observable}, where the step is predicted on extremely small distance scales.

We have discussed a twinflationary model of double inflation with a break for simplicity, but there is a parametric regime where double inflation without a break gives the required amount of asymmetric reheating into the Standard Model. With two periods of inflation, the second period dilutes the energy density of the heavier field sufficiently that there is no observable signal of it produced in reheating. However, even with only one period, inflation can continue for long enough after the inflaton turns the corner in field space such that, at late times, the fraction of the inflaton in the $B$ state relative to the $A$ state is small enough that the expected energy densities that are transferred into each sector satisfy $\rho_B/\rho_A < 0.1$. This occurs as long as $r \gtrsim 1.2$, assuming that the mixing angle of the slow-rolling field with the $\varphi_A$ and $\varphi_B$ fields entirely determines the fraction of its energy that reheats each sector. There is thus a much larger range of $r$ where this toy model of inflation passes $N_{\rm eff}$ bounds than our above analysis shows. The resulting imprint on the CMB could resemble that of the long-lived decay model of Section \ref{sec:late}, with $\Delta N_{\rm eff}$ again being related to the ratio of branching fractions, although this is dependent upon the UV completion of the Twin Higgs.


When there is only one period of inflation, the step is smoothed out and less pronounced and it is necessary to locate the feature numerically. Furthermore, having multiple degrees of freedom available allows for non-trivial evolution of momentum modes after they become super-horizon, which does not occur in single-field inflation but may be calculated from the full solution to the field equations \cite{Wands:2002bn}. While a twinned potential leading to two periods of inflation generally predicts a step in the power spectrum, when there is no break the predictions, and thus constraints, this prediction become more model-dependent. Therefore we leave detailed predictions in that case for future study using realistic models and merely state that the range of $r = 1$ to $n$ interpolates between the single field spectrum and that with a step, as one would expect. 

There are also at least two other detectable effects one might expect in double inflation without a break and in general realistic twinflationary models. Interactions between inflaton fields may produce primordial non-Gaussianities, while the presence of additional oscillating degrees of freedom may produce isocurvature perturbations. These do not appear in our toy model because the heavy field is exponentially damped during the second inflation. CMB-S4 is projected to improve Planck's bounds on non-Gaussianities by a factor of $\sim 2$ and on isocurvature perturbations by perhaps an order of magnitude (though model-independent projections have not been made), so may be able to detect or place useful constraints on realistic twinflationary models \cite{CMB-S4:2016}.

We have introduced twinflation as a mirror Twin Higgs model which suppresses the cosmological effects of twin light degrees of freedom. It extends the mirror symmetry to the inflationary sector. The soft $Z_2$ symmetry-breaking of the Higgs sector may be used in the inflationary sector to cause distinct periods of inflation. There exists a parametric region where this is cosmologically indistinct from single-stage inflation, but also another in which it may be observable. As the direct product of inflation and the Mirror Twin Higgs, this is in some sense a minimal solution.

\section{Conclusion} \label{sec:conc}

In this work we have considered scenarios in which cosmology provides meaningful insight on solutions to the electroweak hierarchy problem. In particular, we have demonstrated several simple mechanisms in which the cosmological history of a mirror Twin Higgs model is reconciled with current CMB constraints and provides signatures accessible in future CMB experiments. In the case of out-of-equilibrium decays, we have found that decays of $\mathbb{Z}_2$-even scalars sufficiently dilute the energy density in the twin sector without the addition of any new sources of $\mathbb{Z}_2$-breaking. In much of the parameter space, the residual contribution to $\Delta N_{\rm eff}$ is directly proportional to the ratio of vacuum expectation values $v^2/f^2$ parameterizing the mixing between Standard Model and twin sectors (as well as the tuning of the electroweak scale), and may be within reach of CMB-S4 experiments. In the case of twinflation, we have found that a (broken) $\mathbb{Z}_2$-symmetric inflationary sector may successfully dilute the energy density in the twin sector, as well as potentially leave signatures in the form of a step in the primordial power spectrum or in departures of primordial perturbations from adiabaticity and Gaussianity. In both cases, these models raise the tantalizing possibility that signatures of electroweak naturalness may first emerge in the CMB, rather than the LHC.

There are a variety of possible directions for future work. Here we have focused on the cosmological consequences of late-decaying scalars and twinned inflationary sectors without specifying their origin in a microscopic model. It would be interesting to construct complete models (where, e.g., supersymmetry or compositeness protect the scale $f$ from UV contributions) in which the existence and couplings of late-decaying scalars arise as intrinsic ingredients of the UV completion. Likewise, we have considered only a toy model of twin chaotic inflation; it would be interesting to see if twinflation may be realized in complete inflationary models that match the observed spectral index and constraints on the tensor-to-scalar ratio.

While we have taken care to ensure that our scenarios respect the well-measured cosmological history beneath $T \sim$ 1 MeV, we have not addressed the origin of the observed baryon asymmetry. In the case of out-of equilibrium decays, there are a number of possibilities. It is plausible that a somewhat larger baryon asymmetry is generated through various conventional mechanisms and diluted by late decays. Alternatively, the decay mechanism itself may possibly be expanded to generate a baryon asymmetry or some other late decay may generate the baryon asymmetry below $\sim 1$ GeV. In the case of twinflation, inflationary dilution of pre-existing baryon asymmetry requires that baryogenesis occur in association with reheating or via another mechanism at temperatures below $\sim 1 \ \text{GeV}$. It would be worthwhile to study models for the baryon asymmetry consistent with these scenarios. Steps in this direction have been taken in \cite{Farina:2016ndq}, which attempted to relate this to asymmetric dark matter in the twin sector.

Likewise, any investigation of dark matter, be it related directly to the twin mechanism or otherwise, must also address implications of the dilution. Previous work attempting to construct dark matter candidates in the twin sector \cite{Barbieri:2016zxn,Garcia:2015loa,Craig:2015xla, Garcia:2015toa, Farina:2015uea, Freytsis:2016dgf,Farina:2016ndq, Prilepina:2016rlq}) has relied upon explicit $\mathbb{Z}_2$-breaking that is not present in the mirror model. Dark matter may alternatively be unrelated to the Twin Higgs mechanism, such as a a WIMP in some minimal extension of the electroweak sector that freezes-out as an overabundant thermal relic and is then diluted to the observed density during reheating. Alternatively, it may be that the dark matter abundance is produced directly during reheating. It would be interesting to study extensions of our scenarios that incorporate dark matter candidates directly related to the mechanism of dilution.

Finally, we have only approximately parameterized Planck constraints and the reach of CMB-S4 on twin neutrinos and twin photons. Ultimately, more precise constraints and forecasts may be obtained via numerical CMB codes. This strongly motivates the future study of CMB constraints on scenarios with three sterile neutrinos and additional dark radiation whose temperatures differ from the Standard Model thermal bath.

\acknowledgments
We thank Zackaria Chacko, Yanou Cui, Paddy Fox, Daniel Green, and Roni Harnik for useful discussions, and Nima Arkani-Hamed for enlightening remarks on $N$naturalness. The work of NC, SK, and TT is supported in part by the US Department of Energy under the grant DE-SC0014129. NC acknowledges the hospitality of the Kavli Institute for Theoretical Physics and the corresponding support of the National Science Foundation under Grant No. NSF PHY11-25915.


\bibliography{twincosmorefs}
\bibliographystyle{JHEP}

\end{document}